\documentclass[10pt, pre, amsmath, onecolumn, showpacs, superscriptaddress]{revtex4-1}
\usepackage{graphicx,color}
 \graphicspath{{./}{./figures/}}
\usepackage{ams math}
\usepackage{enumerate}
\usepackage{mathbbol}
\usepackage{amsfonts}
\usepackage{natbib}

\usepackage{bm}
\usepackage{bigints}
\usepackage{relsize}

\usepackage{color}
\usepackage[dvipsnames]{xcolor}

\usepackage{bbm}

\newcommand{\be}{\begin{equation}}
\newcommand{\ee}{\end{equation}}
\newcommand{\bea}{\begin{eqnarray}}
\newcommand{\eea}{\end{eqnarray}}
\newcommand{\beq}{\begin{eqnarray}}
\newcommand{\eeq}{\end{eqnarray}}

\usepackage[utf8]{inputenc}

\usepackage{amsmath} %math stuff
\usepackage{amsthm} %proof environment
\usepackage{amssymb} %for some math symbols
\usepackage{dsfont} %for identity 1
\usepackage[english]{babel}
\usepackage{floatrow} %to insert rightside caption
\usepackage{graphicx} 
\usepackage{hyperref} %to create links to equations or sections
\usepackage{mathtools}  %for colonneq and other commands

\DeclareMathOperator{\erf}{erf}

\begin{document}

\title{Current fluctuations  in stochastically resetting particle systems}

\author{Costantino \surname{Di Bello}}
\affiliation{Institute for Physics \& Astronomy, University of Potsdam, 14476 Potsdam-Golm, Germany}
\author{Alexander K. \surname{Hartmann}}
\affiliation{Institut f\"ur Physik, Universit\"at Oldenburg, Oldenburg, Germany}
\author{Satya N.~\surname{Majumdar}}
\affiliation{LPTMS, CNRS, Univ. Paris-Sud, Universit\'e Paris-Saclay, 91405 Orsay, France}
\author{Francesco \surname{Mori}}
\affiliation{Rudolf Peierls Centre for Theoretical Physics, University of Oxford, Oxford, United Kingdom}
\author{Alberto \surname{Rosso}}
\affiliation{LPTMS, CNRS, Univ. Paris-Sud, Universit\'e Paris-Saclay, 91405 Orsay, France}
\author{Gr\'egory \surname{Schehr}}
\affiliation{Sorbonne Universit\'e, Laboratoire de Physique Th\'eorique et Hautes Energies, CNRS UMR 7589, 4 Place Jussieu, 75252 Paris Cedex 05, France}

%\date{November 2022}

\begin{abstract}
We consider a system of non-interacting particles on a line with initial positions distributed uniformly with density $\rho$ on the negative
half-line. We consider two different models: (i) each particle performs independent Brownian motion with stochastic resetting to its initial position
with rate $r$ and (ii) each particle performs run and tumble motion, and with rate $r$ its position gets reset to its initial value and simultaneously its velocity gets randomised. We study the effects of resetting on the distribution $P(Q,t)$ of the integrated particle current $Q$ up to time $t$ through the origin (from left to right). We study both the annealed and the quenched current distributions and in both cases, we find that resetting induces a stationary limiting distribution of the current at long times. However, we show that the approach to the stationary state of the current distribution in the annealed and the quenched cases are drastically different for both models. In the annealed case, the whole distribution $P_{\rm an}(Q,t)$ 
approaches its stationary limit uniformly for all $Q$. In contrast, the quenched distribution $P_{\rm qu}(Q,t)$ attains its stationary form for 
$Q<Q_{\rm crit}(t)$, while it remains time-dependent for $Q > Q_{\rm crit}(t)$. We show that $Q_{\rm crit}(t)$ increases linearly with $t$ for large $t$. On the scale where $Q \sim Q_{\rm crit}(t)$, we show that $P_{\rm qu}(Q,t)$ has an unusual large deviation form with a rate function that has a third-order phase transition at the critical point. We have computed the associated rate functions analytically for both models. Using an importance sampling method that allows to probe probabilities as tiny as $10^{-14000}$, we were able to compute numerically this non-analytic rate function for the resetting Brownian dynamics and found excellent agreement with our analytical prediction. 
\end{abstract}

\maketitle

\section{Introduction}

In a non-equilibrium open system, where there is typically a flow of particles or energy from one region of space to another, one of the central observables is the current fluctuation at a fixed point in space~\cite{prahofer-spohn-2002,bodineau-derrida-prl2004,bertini-landim-prl2005,bertini-jstatphys2006,prolhac-mallick-08,derrida-lecomte-wijland-pre2008,derrida-gers,derrida-gers-sep,KM12,SD15,SD16,BJC22,JDR23,BKP2019,Banerjee_2020,MB22,Marbach21}. For example, in a one-dimensional setting, one measures the current $Q(t)$ denoting the number of particles that have passed through a fixed point in space (say from left to right) up to a fixed time $t$, starting from a given initial condition. The current $Q(t)$ of course is a random variable with a distribution $P(Q,t) = {\rm Prob.}(Q(t)=Q)$ and it has two sources of randomness~\cite{derrida-gers,derrida-gers-sep}: (i) from the noise-dependence of the trajectories up to time $t$ and (ii) from the randomness of the initial condition. If the distribution of $Q(t)$ is averaged over both sources of randomness simultaneously, this situation is referred to as the ``annealed case'' (in analogy with disordered systems). This gives us the information about the average fluctuations of $Q(t)$ in the system, due to the randomness in the initial conditions. However, the typical fluctuations of $Q(t)$ due to the initial conditions are not captured by the annealed average. To extract these typical fluctuations, one needs to perform a ``quenched average'', as discussed in detail later. As in the case of disordered systems, these two procedures give quite different answers for the current fluctuations. Derrida and Gerschenfeld studied these current fluctuations both for the non-interacting diffusive (Brownian) particles and the symmetric simple exclusion process in one-dimension \cite{derrida-gers,derrida-gers-sep}. They considered the step initial condition, where particles are uniformly distributed to the left of the origin with a uniform density $\rho >0$, while the right of the origin is empty. In this setting, they were able to calculate both the annealed and the quenched current distribution.  

More recently, these results were generalized to other non-interacting particles in one-dimension undergoing a variety of stochastic processes \cite{Banerjee_2020,BJC22}, in particular, for independent run-and-tumble particles (RTPs). The RTP dynamics has generated much current interest in the context of active systems~\cite{Kac74,TC_2008,Berg_book,Martens2012,Weiss,Fodor17,bechinger_active_2016,cates_motility-induced_2015,Cates_Nature,SEB_16,ad-sm-gh,HP95,Malakar18,Mallmin_18,maes,gradenigo,pierre-satya-greg,Bres20}. The dynamics of a  single RTP in one dimension consists of alternating runs and tumbles. During a run, the particle moves ballistically with a constant velocity $+v_0$ during a run time $\tau$ with exponential distribution $p(\tau) = \gamma e^{- \gamma \tau}$ where $\gamma^{-1} >0$ is the persistence time. At the end of the run, the particle tumbles instantaneously, i.e. changes its velocity from $+v_0$ to $-v_0$ and then a new run starts. Thus $v_0$ and $\gamma$ are the two parameters in this model. We consider $N$ independent RTPs, with initial positions distributed independently and uniformly with density $\rho$ on the negative semi-infinite axis (the same step initial condition mentioned above for the diffusive dynamics). Furthermore, we choose the initial velocities to be $\pm v_0$ with equal probability, independently for each particle.  

For a general non-interacting particle systems in one-dimension, starting from the initial conditions mentioned above, it was shown in 
Ref. \cite{Banerjee_2020} that in the annealed case the distribution of $Q(t)$ is universally Poissonian, i.e., 
 \bea \label{Pann:intro}
 P_{\rm an}(Q,t) = e^{-\mu(t)} \frac{\left[\mu(t)\right]^Q}{Q!}\;, \; Q=0,1,2, \cdots \;,
 \eea 
where only $\mu(t)$ depends on the underlying process. In particular this result (\ref{Pann:intro}) admits a large deviation form
\begin{eqnarray}\label{Poisson_largedev}
P_{\rm an}(Q,t) \sim \exp{\left[- \mu(t) \, \Psi_{\rm an} \left( \frac{Q}{\mu(t)}\right) \right]} \;,
\end{eqnarray}  
valid for large $Q$ and $\mu(t)$ with $Q/\mu(t)$ fixed. Here the rate function $\Psi_{\rm an}(q)$ is universal, i.e., independent of the particle dynamics, and is given by
\begin{eqnarray} \label{psi_an}
\Psi_{\rm an}(q) = q \,\ln q - q + 1 \;, \; q \geq 0 \;.
\end{eqnarray}
For the quenched case, the calculation is more intricate. For the non-interacting diffusive case, it was shown in Ref. \cite{derrida-gers,Banerjee_2020} that, for large $t$, large $Q$ with $Q/\sqrt{t}$ fixed, $P{\rm qu}(Q,t)$ admits a large deviation form
 \begin{equation}\label{P-F-diff-largeQ}
P_{\rm qu}(Q,t) \sim \exp\left[-\rho \sqrt{D t} ~ \Psi_{\rm diff}\left(\frac{Q}{\rho \sqrt{D t}}\right)\right] \;,
\end{equation}
where $\Psi_{\rm diff}(q)$ is not fully explicit and grows anomalously as $\Psi_{\rm diff}(q) \sim q^3/12$ as $q \to \infty$~\cite{derrida-gers,Banerjee_2020}. In contrast, for the RTP case, it was shown in \cite{Banerjee_2020} that, for large $Q$ and large $t$ with $Q/t$ fixed, the quenched current distribution takes a large deviation form 
\begin{equation}\label{ac-sig}
P_{\rm qu}(Q,t) \sim \exp\left[-\rho \, v_0 \, \gamma \, t^2 \, \Psi_{\rm RTP}\left(\frac{Q}{\rho \, v_0 \,t}\right)\right],
\end{equation}
where the rate function $\Psi_{\rm RTP}(q)$ is now fully explicit and given by \cite{Banerjee_2020}
\begin{equation}\label{rtp-ldf-model}
\Psi_{\rm RTP}(q)=q-\frac{q}{2}\sqrt{1-q^2}-{\rm sin}^{-1}\left[ \sqrt{\frac{1-\sqrt{1-q^2}}{2}} \right]\;, \quad 0 \leq q \leq 1 \;.
\end{equation}
The upper bound $q \leq 1$ arises from the fact that the total current $Q(t)$ cannot exceed $\rho v_0 t$. In particular, the rate function approaches a nontrivial constant $\Psi_{\rm RTP}(q) \to 1-\pi/4$ as $q\to 1$ \cite{Banerjee_2020}.

Recently, there has been enormous interest in stochastic resetting (for recent reviews see \cite{EMS20,GJ22,PKR22}). Under stochastic resetting, the natural dynamics of a particle gets interrupted at random times and its motion restarts from the initial position. Generically, this resetting drives the system into a non-equilibrium stationary state where the detailed balance is manifestly violated due to the resetting moves and the stationary state carried a nonzero probability current~\cite{EM1,EM2}. In addition, resetting also has rather drastic effects on first-passage properties, but we will not be concerned in this paper with first-passage aspects, but rather with the steady state properties of the system under resetting~\cite{EM1,EM2,MV13,GMS14,MSS15a,Meylahn15,MC16,EM16,MMV17,HT17,EM18,PDRK19,DH2019,MMS20,MR21,SM22,EMS20,GJ22,PKR22,Biroli22,SMS23,FPSRR20,BBPMC20,FBPCM21,VOM22,Ber22,Metzler20,Sandev22}. In particular, our goal is to study the effect of stochastic resetting on the current distribution in a many-body system. One would anticipate that the current distribution also may become stationary at long times under resetting. 

In this paper, we indeed show that a stationary current distribution emerges even for non-interacting particles (both diffusive as well as RTP) in one-dimension undergoing independent stochastic resetting with a constant rate $r$.
For the symmetric exclusion process in the presence of resetting and starting from a step initial condition, the current fluctuation in the annealed case was studied recently~\cite{BKP2019,MB22}. Here we restrict ourselves to noninteracting particles but study the current distribution both in the annealed and in the quenched case. In particular, for the quenched case, the current distribution turns out to be highly nontrivial even for noninteracting particles. At variance with the case without resetting, the current distribution reaches a stationary form in the long time limit due to the resetting dynamics. However, the approach in time of the current distribution to its stationary form is drastically different in the annealed and in the quenched cases, for both diffusive and RTP dynamics. In the annealed case, the whole distribution $P_{\rm an}(Q,t)$ approaches its stationary limit at late times uniformly for all $Q$. In contrast, the quenched current distribution $P_{\rm qu}(Q,t)$ approaches its stationary form in a non-uniform way. More precisely, we show that there exists a critical value $Q_{\rm crit}(t)$ that increases linearly with $t$. 
For $Q < Q_{\rm crit}(t)$, the quenched distribution $P_{\rm qu}(Q,t)$ becomes independent of $t$, while it is still time-dependent for $Q>Q_{\rm crit}(t)$. This critical current $Q_{\rm crit}(t)$ acts as a separatrix between the stationary and the transient regimes in the quenched case. 
In addition, for both dynamics in the quenched case, we find that this transition manifests itself in the large deviation form of $P_{\rm qu}(Q,t) \sim e^{-t^2 \Psi(Q/Q_{\rm crit}(t))}$ when $Q \sim Q_{\rm crit}(t)$. We compute the rate function $\Psi(q)$ analytically for both dynamics and find that it has a third-order singularity at a critical value $q=q_{\rm crit}$ where the rate function and its first two derivatives are continuous but the third derivative is discontinuous. 

Measuring numerically such a rate function is a formidable technical challenge, which we also address in this paper by using an importance sampling algorithm capable of accessing very small probabilities. Using this method we compute numerically the rate function $\Psi(q)$ for the Brownian motion with resetting and find remarkably good agreement with our analytical prediction. Finally, while the large deviations of resetting systems have been studied extensively in the recent past~\cite{Meylahn15,HT17,DH2019,SM22,Bres20}, to the best of our knowledge the one observed in this paper is the first instance of a third-order phase transition for processes with stochastic resetting

The rest of the paper is organised as follows. In Section \ref{sec:model}, we define precisely the two models (diffusive and the RTP) with stochastic resetting and we summarise our main results. In Section \ref{sec:setup}, we recall the general setup to calculate the annealed and the quenched current distributions. In Section \ref{sec:rbm}, we discuss in detail the current distribution for the Brownian particles with stochastic resetting.  
Section \ref{sec:RTP} deals with the current distribution for the run and tumble particles with stochastic resetting. Details of numerical simulations for the Brownian case are presented in Section \ref{sec:num}. Finally, we conclude in Section \ref{sec:conclusion} with a summary and perspectives.

%Among the very wide field of stochastic processes the ones with resetting are the really new star of the last decade. Since the paper of \cite{diffusion_stochastic_resetting}, many researchers devoted their efforts to deepening the understanding of these kind of processes. In particular with `stochastic resetting' we mean that at random times sampled from some distribution, the random walker is instantaneously restored to a specific position in space.

\section{The model and the main results}\label{sec:model}

\subsection{The model}

\begin{figure}
\begin{center}
    \includegraphics[angle=-90,width=0.6\textwidth]{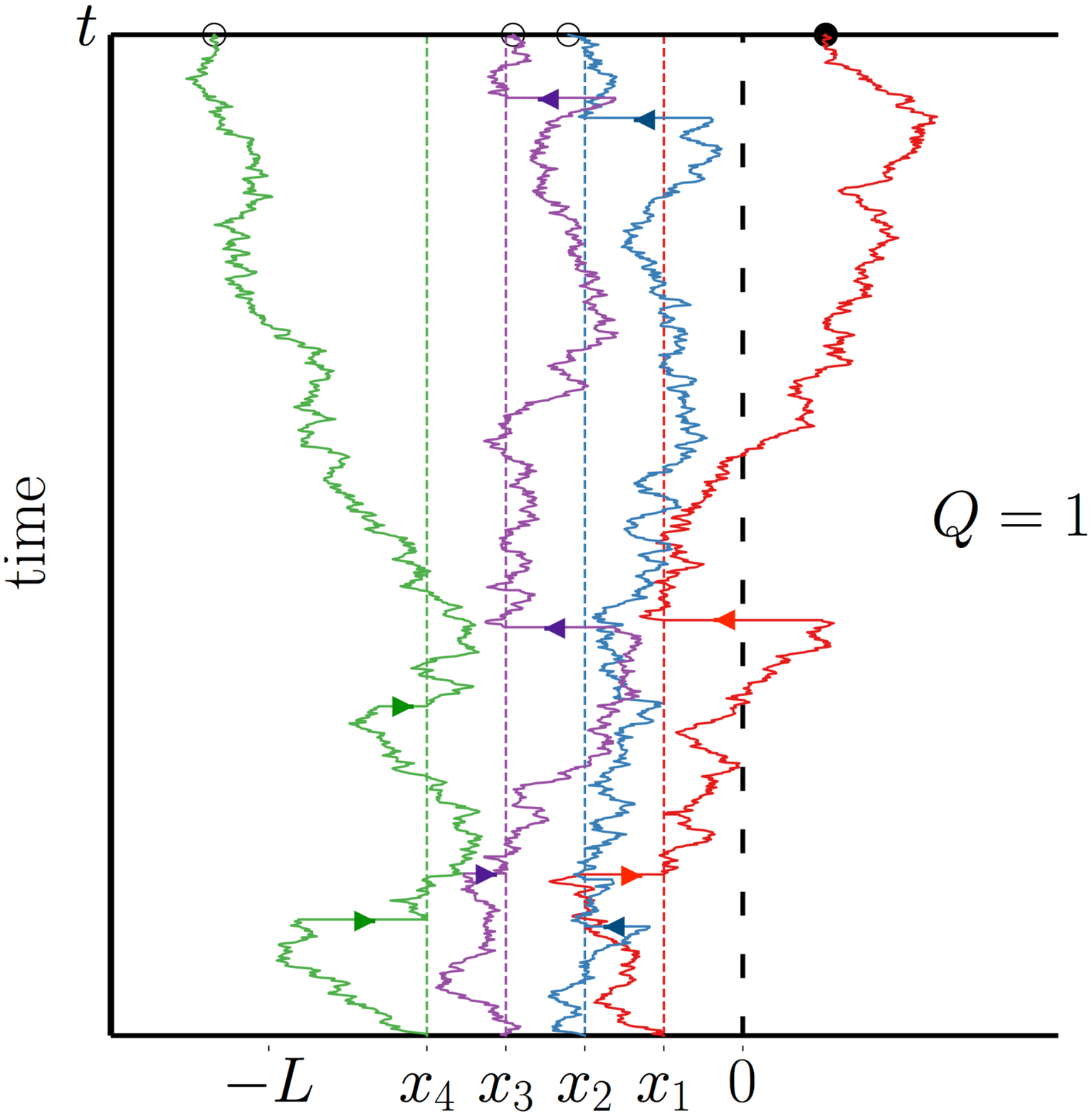}
\end{center}
\caption{\label{fig:system_example} A typical realisation of the trajectories of $N=4$ particles, each undergoing diffusion with stochastic resetting to the initial positions $x_1, x_2, x_3$ and $x_4$. The vertical dashed lines indicate the origin of each trajectory. The quantity $Q = 1$ indicate the number of trajectories that have crossed the red dashed vertical line at position $0$ from left to right up to time $t$. In this example, only the second trajectory, starting from $x_2$, has crossed $0$ up to time $t$. The horizontal arrows in the trajectories represent the resetting events.}
\end{figure}

In this paper we would like to understand how the presence of a stochastic resetting affects the current fluctuations in a system of non-interacting particles. Our model is defined as follows:
\begin{itemize}
    \item there are $N$ non-interacting particles on a line all moving with the same stochastic dynamics. We considered two specific stochastic dynamics
    \begin{itemize}
        \item Brownian motion: here the position of the $i$-th particle evolves via the Langevin equation
        \bea \label{def_BM}
        \frac{dx_i}{dt} = \eta_i(t) \;,
         \eea
         starting from $x_i(0) = x_i$. Here $\eta_i(t)$'s are independent Gaussian white noises with mean $\langle \eta_i(t) \rangle =0$ and correlator $\langle \eta_i(t) \eta_j(t') \rangle= 2D \delta(t-t')\,\delta_{ij}$ with $D$ denoting the diffusion constant.
         
        \item Run-and-tumble particles (RTP): here the position evolves via
        \bea \label{def_RTP}
        \frac{dx_i}{dt} = v_0\, \sigma_i(t) \;,
        \eea
    \end{itemize}
    starting from $x_i(0) = x_i$. Here $v_0$ is the intrinsic speed during a run and $\sigma_i(t)=\pm 1$ are independent (from particle to particle) dichotomous telegraphic noises \cite{Kac74,Weiss}. Each $\sigma_i(t)$ flips from one state to another with a constant rate $\gamma$. The effective noise $\xi_i(t)=v_0\, \sigma_i(t)$ is coloured with an autocorrelation function
\begin{equation}
\langle \xi_i(t) \xi_j(t')\rangle= v_0^2\, e^{-2\, \gamma\, |t-t'|}\,\delta_{ij} \,.
\label{autocorr.1}
\end{equation}
The time scale $\gamma^{-1}$ is the `persistence' time of a run that encodes the memory
of the noise. In the limit $\gamma\to \infty$, 
$v_0\to \infty$ but keeping the ratio $D_{\rm eff}= v_0^2/{2\gamma}$ fixed, the two-time correlator of the effective noises $\xi_i(t)$'s becomes
\begin{equation}
\langle \xi_i(t) \xi_j(t')\rangle= \frac{v_0^2}{\gamma}\, \left[\gamma\, e^{-2\gamma|t-t'|}\right]\, \delta_{ij}
\to 2D_{\rm eff}\, \delta(t-t')\,\delta_{ij} .
\label{autocorr.2}
\end{equation}

    \item the initial positions $x_i$'s of the particles are random variables themselves uniformly distributed in the interval $[-L,0]$. Thus we have an initial step profile for the density $\rho(x)=\rho=\frac{N}{L} \text{ for } x\in [-L,0]$;
    \item each particle is subjected to a Poissonian resetting dynamics with the constant rate $r$ to its initial position. This means that, within a small time interval $dt$, with probability $r \,dt$, the current position $x_i(t) \to x_i$ and with the complementary probability $1-r\,dt$, the current position $x_i(t)$ evolves freely as in Eq. (\ref{def_BM}) or (\ref{def_RTP}) -- see Fig. \ref{fig:system_example} for typical trajectories for the diffusive case with resetting. Note that in the case of the RTP, one has to be more careful in defining the resetting protocol. Here, following Ref. \cite{EM18},  we start each RTP with a random velocity
$\pm v_0$ with equal probability. When the position $x_i(t)$ is reset to the initial value $x_i$, we assume that the velocity
gets randomised and the process (both position and velocity) renews itself after each resetting.

\end{itemize}

\subsection{The observable of interest: annealed vs. quenched current distribution}

We start the stochastic dynamics of the system from the initial positions $\{x_1, x_2, \cdots, x_N \}$ and evolve the system up to time $t$.
Let $Q(t)$ denote the total current up to time $t$, i.e., the number of particles that have crossed the origin from left to right up to $t$. It was realized in Ref. \cite{Banerjee_2020} that, for the step initial condition, this history-dependent observable $Q(t)$ is related to an instantaneous observable at time $t$, namely the number of particles $N_+(t)$ present at time $t$ on the right of the origin. The distribution of $Q(t)$ clearly depends on the initial positions and we denote it by $P(Q,t|\{x_1, x_2, \cdots, x_N\})$. It is convenient to consider the generating function of $Q(t)$, namely
\begin{equation} \label{def_GF}
    \langle e^{-pQ}\rangle_{\lbrace x_i \rbrace} \coloneqq \sum_{Q=0}^{N} e^{-pQ}P(Q,t \mid \lbrace x_i\rbrace)
\end{equation}
The annealed and quenched averages are defined as \cite{derrida-gers,Banerjee_2020}
\begin{gather}
    \sum_{Q=0}^{N} e^{-p Q} P_{\mathrm{an}}(Q, t)=\overline{\left\langle e^{-p Q}\right\rangle_{\left\{x_{i}\right\}}} \label{def_ann}\\
   \int_0^\infty e^{-p Q} P_{\mathrm{qu}}(Q, t)\, dQ =\exp \left[\overline{\ln \left\langle e^{-p Q}\right\rangle_{\left\{x_{i}\right\}}}\right] \label{def_qu} \;.
\end{gather}
where the symbol $\overline{(\dots)}$ represents the average over $\lbrace x_i \rbrace$. Note that in the quenched case, by taking first the average of the logarithm followed by re-exponentiation allows to extract the typical behavior of $P(Q,t)$ amongst all possible initial configurations. In the set of typical configurations (e.g., approximately equi-spaced initial positions), the value of $Q$ is rather large and hence the discrete sum in the annealed case gets replaced by an integral over continuous values of $Q$. Hence the generating function gets replaced by the Laplace transform in the quenched case. 
In this paper, we compute exactly both the annealed and the quenched current distributions for the diffusive as well as for the RTP case with stochastic resetting. Our main results are summarised in the next subsection.

\subsection{Main results}\label{sec:main}

\vspace*{0.4cm}
\noindent
{\bf Annealed case}: In this case, it was already shown in Ref. \cite{Banerjee_2020} that, irrespective of the dynamics, the distribution $P_{\rm an}(Q,t)$ has a universal Poissonian form
\begin{equation}\label{Pan-model-def}
P_{\rm an}(Q,t) = e^{-\mu(t)} \frac{\left[\mu(t)\right]^Q}{Q!} \;, \; Q=0,1,2,\cdots \;.
\end{equation}
Only the mean $\mu(t)$ depends on the underlying process explicitly as
\begin{equation}\label{muan-mod-def}
\mu(t) = \rho \, \int_0^{\infty} dz ~U(z,t) \;,
\end{equation} 
where
\begin{equation}
    \label{eqn:U_definition}
    U(z,t) \coloneqq \int_{0}^{+\infty} \mathrm{d}x \, G(x, -z,t), \quad z\geq 0 \;.
\end{equation}
In this expression, $G(x, -z,t)$ is the Green's function of the underlying dynamics, i.e., the probability density, for a single
particle, to reach $x$ at time $t$, starting from $-z$. In this paper, we compute the mean $\mu_r(t)$ (the subscript $r$ refers to resetting with rate $r$), both for the diffusive and the RTP case. For the diffusive case, it is given by
\begin{equation}
    \label{eqn:mu_Brownian}
    \mu_r(t) = \dfrac{\rho}{2} \,\sqrt{\dfrac{D}{r}}\, \erf\left(\sqrt{rt} \right) \;,
\end{equation}
where $\erf(z) = (2/\sqrt{\pi})\,\int_0^z e^{-u^2}\,du$ is the error function. The function $\mu_r(t)$ has the following asymptotic behaviors
\bea \label{mur_asympt_diff}
\mu_r(t) \approx
\begin{cases}
& \rho \sqrt{\frac{D}{\pi}}\, \sqrt{t} \quad, \quad t \to 0 \;,\\
& \\
& \frac{\rho}{2} \, \sqrt{\frac{D}{r}}\quad, \quad \quad t \to \infty \;.\\
\end{cases}
\eea 
Thus, as time increases, the mean current initially increases as $\sqrt{t}$, as in the case without resetting and saturates exponentially fast to its
stationary value as $t \to \infty$. 

In the case of the RTP, we show that 
\begin{equation}
    \label{eqn:renewal_mu}
    \mu_r(t) = e^{-rt}\mu_0(t) +r\int_0^t e^{-r\tau}\mu_0(\tau) \mathrm{d}\tau \;,
\end{equation}
where $\mu_0(t)$ is the result for RTP without resetting found in \cite{Banerjee_2020}
\begin{equation}
    \mu_0(t) = \dfrac{1}{2} \rho v_0 te^{-\gamma t} [I_0(\gamma t) +I_1(\gamma t)] \;,
\end{equation}
with $I_0,I_1$ denoting the modified Bessel functions of the first kind. The asymptotic behaviors of $\mu_r(t)$ are given by
\bea \label{mur_asympt_RTP}
\mu_r(t) \approx
\begin{cases}
&\frac{1}{2} \rho\, v_0\, t \quad, \quad \quad \; \,t \to 0 \;, \\
& \\
& \frac{\rho \,v_0}{2 \sqrt{r(r+2 \gamma)}} \quad, \quad t \to \infty \;.
\end{cases}
\eea
As in the case of resetting Brownian motion, the mean current again approaches exponentially fast to a stationary limiting value as $t \to \infty$. 
In the diffusive limit $v_0 \to \infty$, $\gamma \to \infty$ with $v_0^2/(2\gamma) = D$ fixed, the stationary value in the second line of Eq. (\ref{mur_asympt_RTP}) coincides with the stationary value for the resetting Brownian motion given in Eq.  (\ref{mur_asympt_diff}).

Thus in the long time limit, in both models, the annealed current distribution in Eq. (\ref{Pan-model-def}) converges to a stationary distribution
\bea \label{P_an_stat_intro}
P_{\rm an}(Q,t) \underset{t\to \infty}{\longrightarrow} e^{-\mu_r(\infty)} \frac{\left[\mu_r(\infty)\right]^Q}{Q!} \;, \; Q=0,1,2,\cdots \;.
\eea
This stationary distribution is still Poissonian but with a constant mean $\mu_r(\infty)$ as given in Eqs. (\ref{mur_asympt_diff}) and (\ref{mur_asympt_RTP}) for the two models. Thus, for any value of $Q$, the distribution $P_{\rm an}(Q,t)$ converges uniformly to its stationary value exponentially fast.

\vspace*{0.4cm}
\noindent
{\bf Quenched case}: As in the annealed case, in the limit $t \to \infty$, the quenched current distribution $P_{\rm qu}(Q,t)$ also approaches a stationary distribution in the presence of resetting in both models. However, the approach to the stationary state in the quenched case is very different from that of the annealed case. We recall that the annealed distribution approaches its stationary form uniformly for all $Q$. In contrast, 
for the quenched case, we show that there is a critical value $Q_{\rm crit}(t)$ that increases linearly with $t$ such that, for $Q < Q_{\rm crit}(t)$ the quenched distribution attains its stationary form, while it is still time-dependent for $Q > Q_{\rm crit}(t)$. Thus the stationary state gets established on longer and longer scales as time increases. The critical value $Q_{\rm crit}(t)$ separates the steady state and the transient regime. This is reminiscent of how the position distribution evolves for a single resetting Brownian motion studied in Ref. \cite{MSS15a}. On this scale, when $Q \sim Q_{\rm crit}(t)$, we show that the quenched
distribution admits an unusual large deviation form in both models, that exhibits a third order phase transition at $Q = Q_{\rm crit}(t)$. 

For the diffusive case, we find that, in the limit $Q \to \infty$, $t \to \infty$ keeping the ratio $q = Q/(rt)$ fixed, the quenched current distribution admits a large deviation form 
\begin{equation}
    P_{\mathrm{qu}}(Q,t) \sim \exp \left[-r^2 t^2 \Psi^{(r)}_{\mathrm{diff}}\left(\dfrac{Q}{rt}\right) \right] \;, \label{Pqu_diff}
\end{equation}
with 
\begin{equation}
\label{large_deviation_function_psi_diffusion}
\Psi^{(r)}_\mathrm{diff}(q)=\begin{cases}
\dfrac{q^2}{q^*}\quad & \text{ for } \quad q<q^*=2\sqrt{\dfrac{D}{r}}\,\rho\\
-\dfrac{q^*}{3}+q+\dfrac{q^3}{3{q^*}^2} & \text{ for } \quad q>q^*=2\sqrt{\dfrac{D}{r}}\,\rho \;.
\end{cases}
\end{equation}
We note that the function $\Psi^{(r)}_{\mathrm{diff}}(q)$ and its two first derivatives are continuous at $q=q^*$, while the third derivative is discontinuous, indicating a third order dynamical phase transition. In fact, taking the limit $t \to \infty$ but keeping $Q$ fixed, one sees, using the result in the first line of Eq. (\ref{large_deviation_function_psi_diffusion}), that the current distribution $P_{\rm qu}(Q,t)$ approaches a stationary Gaussian form
\bea \label{P_qu_diff_stat}
P_{\rm qu}(Q,t) \underset{t \to \infty}{\longrightarrow} \exp{\left(-\frac{1}{2\rho}\,\sqrt{\frac{r}{D}}\,Q^2\right)} \;.
\eea 
Indeed, for finite but large $t$, for $Q \ll q^* r\,t$, the current distribution becomes time-independent, while for $Q \gg q^* r\,t$, the
distribution is transient, i.e., time-dependent. Thus the boundary $Q_{\rm crit}(t) = q^*r\,t$ in the $(Q,t)$-plane separates the stationary regime from the 
transient regime discussed above, indicating the two phases across the third order phase transition at $q = q^*$.

In the case of RTP, we show that the quenched current distribution $P_{\mathrm{qu}}(Q,t)$ similarly admits 
a large deviation form in the limit $Q \to \infty$, $t \to \infty$ but keeping $q = Q/(\rho\, v_0\, t)$ fixed
\begin{equation} \label{large_dev_qu_intro}
    P_{\mathrm{qu}}(Q,t) \sim \exp \left[ -\rho v_0 \gamma\,t^2 \, \Psi^{(r)}_{\mathrm{RTP}}\left( \dfrac{Q}{\rho v_0 t} \right) \right] \;,
\end{equation}
where
\bea \label{Psi_r_intro}
    \Psi^{(r)}_{\mathrm{RTP}}(q)=
    \begin{cases}
&\dfrac{q_c}{2\sqrt{1 - q_c^2}}\, q^2 \quad, \quad \hspace*{7.8cm} 0< q < q_c = \dfrac{\sqrt{\alpha(\alpha+2)}}{\alpha+1}\;, \\
& \\
& -\dfrac{q_c}{2 \sqrt{1- q_c^2}} + \dfrac{q}{\sqrt{1-q_c^2}} - \dfrac{1}{2} q\,\sqrt{1 - q^2} + \dfrac{1}{2} {\rm sin}^{-1}(q_c) - \dfrac{1}{2} {\rm sin}^{-1}(q) \quad, \quad q_c < q \leq 1 \;.
\end{cases}
   \eea
with $\alpha = r/\gamma$. This rate function also undergoes a third-order phase transition at $q=q_c$, similar to the resetting Brownian motion.   
In this case, $Q_{\rm crit}(t) = \rho\,v_0\, q_c\,t$ separates the steady state regime for $Q < Q_{\rm crit}(t)$ and the transient regime for $Q > Q_{\rm crit}(t)$. 
Note that in the limit $r\to 0^+$ (or $\alpha \to 0^+$), when $q_c \to 0$, we recover the result in Eq. (\ref{rtp-ldf-model}) without resetting.

\section{General setup for non-interacting particles: current fluctuations}\label{sec:setup}

The general setup for the current fluctuations for non-interacting particles starting from a step initial condition was already
discussed in Ref. \cite{Banerjee_2020}. We briefly recall this setup here to make the discussion in this paper self-contained and in addition, this will also serve the purpose of setting our notations. Below we discuss separately the annealed
and the quenched cases.

\subsection{Annealed case}\label{sec:annealed}

The current distribution in the annealed case for any stochastic dynamics of non-interacting particles, starting from the step initial condition, was shown to be Poissonian with mean $\mu(t)$ ant it can be expressed in terms of the Green's function of the underlying dynamics. The main observation of Ref. \cite{Banerjee_2020} was that the total current $Q(t)$ (left to right through the origin) up to time $t$ coincides with the number of particles $N_+(t)$ at time $t$ that are present on the right of the origin, i.e.  
\begin{equation}\label{N+defn}
 Q(t) = N_+(t)= \sum_{i=1}^N \theta(x_i(t)) \;,
\end{equation}
where $\theta(x)= 1$ if $x\geq 0$ and zero otherwise. For a fixed initial condition $\{x_1, x_2, \cdots, x_N \}$ it follows that the generating function
\begin{equation}\label{lpbasic}
 \sum_{Q=0}^{\infty} e^{-pQ} P(Q,t,\{x_i\}) = \langle e^{-pQ}\rangle_{\{x_i\}} = \left \langle \exp[{-p \sum_{i=1}^N \theta(x_i(t))}]\right \rangle_{\{x_i\}} = \prod_{i=1}^N\left[1- (1-e^{-p})\langle \theta(x_i(t)) \rangle \right] \;,
 \end{equation}
where we used $e^{-p \theta(x)} = 1 -(1-e^{-p})\theta(x)$ is the last equality and the independence of particles. Noticing that 
\begin{equation}\label{I-propagator}
  \langle \theta(x_i(t)) \rangle = \int_0^{\infty} G(x,x_i,t) dx = U(-x_i,t) \;, \quad x_i < 0 \;,
 \end{equation}
where $G(x,x_i,t)$ is the single-particle Green's function, i.e., the propagator for a particle to reach $x$ at time $t$, starting initially at $x_i<0$. This gives
\begin{equation}\label{propagator}
  \langle e^{-pQ}\rangle_{\{x_i\}}= \prod_{i=1}^N\left[1- (1-e^{-p})U(-x_i,t)\right]  \;, \quad x_i < 0 \;, \quad \forall i = 1, \cdots, N \;.
 \end{equation}
This is a general result, valid for any underlying dynamics, the information on the dynamics being contained in $U(z,t)$. 

The annealed distribution $P_{\rm an}(Q,t)$ is defined in Eq. (\ref{def_ann}). Performing the average $\overline{\cdots}$ over the initial positions 
gives 
\begin{equation}\label{ann-G-avg}
\langle\overline{e^{-pQ}\rangle_{\{x_i\}}}= \prod_{i=1}^N\left[1- (1-e^{-p})\, \overline{U(-x_i,t)}\right] \;,
\end{equation}
where $U(-x_i,t)$ is defined in Eq. (\ref{I-propagator}). To perform the average over the initial conditions with a fixed uniform density $\rho$, 
we assume that each of the $N$ particles is distributed independently and uniformly over a box $[-L,0]$ and then eventually take the limit 
$N \to \infty$, $L \to \infty$ keeping the density $\rho = N/L$ fixed. For this uniform measure, each $x_i$ is uniformly distributed in the box
$[-L,0]$. Using the independence of the $x_i$'s we then get 
\begin{equation}\label{x_i-avg}
\langle\overline{e^{-pQ}\rangle_{\{x_i\}}} = \prod_{i=1}^N \left[1- (1-e^{-p}) \int_{-L}^{0} U(-x_i,t) \frac{dx_i}{L} \right] = \left[1- \frac{1}{L}(1-e^{-p}) \int_0^L U(z,t) dz \right]^N \;,
\end{equation}
where, in the last equality, we made the change of variable $z=-x_i$. Taking now the limit $N \to \infty$, $L \to \infty$ keeping $\rho = N/L$ fixed gives
\begin{eqnarray}\label{lap-pan}
\sum_{Q=0}^{\infty} e^{-pQ} P_{\rm an}(Q,t) = \langle\overline{e^{-pQ}\rangle_{\{x_i\}}} = \exp\left[-\mu(t) ~ (1-e^{-p})\right] \;, \quad {\rm where} \;\quad \mu(t)=  \rho \int_0^{\infty} dz ~ U(z,t) \;.
\end{eqnarray}
By expanding $\exp\left[-\mu(t) ~ (1-e^{-p})\right]$ in powers of $e^{-p}$ and comparing to the left hand side, we see that $Q$ can take only integer values $Q=n=0,1,2,\cdots$ and the probability distribution is thus simply a Poisson distribution with mean $\mu(t)$ given in Eq. (\ref{lap-pan}).
Thus, all we need to compute is the mean $\mu(t)$ for a given process.

\subsection{Quenched case}

As in the annealed case, the general formalism for the quenched case, valid for non-interacting particles  with arbitrary dynamics,
was also worked out in Ref. \cite{Banerjee_2020}. Here, we briefly recall this formalism for the sake of completeness. We start with the definition
of the current distribution in the quenched case in Eq. (\ref{def_qu}), which we recall here
\begin{equation}\label{quenched-definition}
\int_{0}^{\infty} P_{\rm qu} (Q,t) e^{-pQ} \, dQ = \exp \left[\overline{{\ln}\left[\langle e^{-pQ}\rangle_{\{x_i\}} \right]} \right] ,
\end{equation}
where $\overline{\cdots}$ represents an average over the initial positions $\{x_i\}$. Our starting point is again Eq. (\ref{propagator}). We take the logarithm on both sides of (\ref{propagator}) to obtain
\begin{equation}\label{Nlog-quenched}
{\ln}\left[\langle e^{-pQ}\rangle_{\{x_i\}} \right] = \sum_{i=1}^N {\ln}\left[1-(1-e^{-p})U(-x_i,t) \right] \;,
\end{equation}
where $U(-x_i,t)$ is defined in Eq. (\ref{I-propagator}) in terms of the single particle Green's function $G(x,x_i,t)$ of the underlying dynamics. 
Next, we perform the average over the initial positions where each $x_i$ is chosen independently and uniformly from the box $[-L,0]$. 
Taking the thermodynamic limit $N \to \infty$, $L \to \infty$ with $\rho = N/L$ fixed, we get 
\begin{equation}\label{Nlog-qu-avg}
\overline{{\rm log}\left[\langle e^{-pQ}\rangle_{\{x_i\}} \right]}= \frac{N}{L}\int_{-L}^0 dx_i ~{\ln}\left[1-(1-e^{-p})U(-x_i,t) \right] \longrightarrow \rho \int_0^\infty dz\, \ln \left[ 1 - (1-e^{-p}) U(z,t)\right] \;.
\end{equation} 
Therefore the Laplace transform of the quenched flux distribution is given by
\begin{equation}\label{Pqu2}
\int_{0}^{\infty} P_{\rm qu} (Q,t) e^{-pQ}\, dQ  = \exp\left[I(p,t)\right] \;,
\end{equation}
where 
\begin{eqnarray}\label{def_Ip}
I(p,t) = \rho \int_0^\infty dz\, \ln \left[ 1 - (1-e^{-p}) U(z,t)\right] \;.
\end{eqnarray}
This result is very general and holds for any stochastic process. While it is difficult to invert this Laplace transform (\ref{Pqu2}) and (\ref{def_Ip}) to get $P_{\rm qu} (Q,t)$ explicitly, one can easily derive its moments by making a small $p$ expansion of (\ref{def_Ip}). For example, for the mean and the variance one gets \cite{Banerjee_2020}
\begin{eqnarray}
&&\langle Q \rangle_{\rm qu} = \rho \, \int_0^\infty U(z,t) \, dz \;,  \label{mean_qu} \\
&&\sigma_{\rm qu}^2 = \langle Q^2 \rangle_{\rm qu} - \langle Q \rangle^2_{\rm qu} = \rho \int_0^\infty U(z,t)(1-U(z,t))\, dz \;. \label{var_qu}
\end{eqnarray}

Thus, to summarise, both in the annealed and the quenched cases, the basic information needed to compute the current distribution is contained 
in the function $U(z,t) = \int_0^\infty G(x,-z,t)\,dx$. This is indeed the central quantity and to compute it, all we need is the single particle
Green's function of the underlying process. Below we derive the results for the diffusive and the RTP dynamics with resetting separately, both for the annealed and the quenched cases. 

\section{Brownian particles with stochastic resetting}\label{sec:rbm}

Here, the underlying dynamics of each particle is a Brownian diffusion with stochastic resetting with rate $r$ to its initial position. The initial positions are distributed independently and uniformly with density $\rho$ on the negative real line. Both for the annealed and the quenched cases, the central quantity needed is the function $U_r(z,t) = \int G_r(x,-z,t)\,dx$ where $G_r(x,-z,t)$ is the propagator of a single resetting Brownian motion from $-z$ to $x$ in time $t$.
Note that the subscript $r$, here, denotes the ``resetting'' Brownian motion. This propagator can be easily derived from a simple renewal equation \cite{EMS20} 
\begin{equation}
    \label{eqn:last_renewal_eq}
    G_r(x,x_0,t) = e^{-rt}G_0(x,x_0,t) +r\int_{0}^{t} e^{-r\tau} G_0(x,x_0, \tau) \mathrm{d}\tau \;,
\end{equation}
where $G_0(x,x_0,\tau) = e^{-(x-x_0)^2/(4 D \tau)}/\sqrt{4 \pi D \tau}$ is just the Brownian propagator without resetting. The result in Eq. (\ref{eqn:last_renewal_eq}) can be understood as follows. First we consider the case where there is no resetting in the interval $[0,t]$, which happens with probability $e^{-rt}$. In this case, the propagator is simply $G_0(x,x_0,t)$, explaining the first term in Eq.~(\ref{eqn:last_renewal_eq}). In case when there are multiple resettings in $[0,t]$ to the initial position $x_0$, it suffices to keep track of what happens after the last resetting before $t$. Let this last resetting happen at time $t-\tau$. Then between $t-\tau$ and $t$, the particle evolves freely from $x_0$ during the interval $\tau$ and hence its propagator is simply $G_0(x,x_0, \tau)$. Furthermore, the factor $r\,e^{-r\tau}\,\mathrm{d}\tau$ denotes the probability of the event that there are no resetting in the interval $[t-\tau,t]$, preceded by a resetting event in the time interval $\mathrm{d}\tau$ before $t-\tau$. Multiplying these two probabilities and integrating over $\tau \in [0,t]$ gives the second term in Eq. (\ref{eqn:last_renewal_eq}). To solve this equation, it is convenient to take the Laplace transform with respect to $t$, which gives
\begin{equation}
    \label{eqn:last_renewal_Laplace}
   \tilde{G}_r(x,x_0,s)= \int_0^\infty  G_r(x,x_0,t) \, e^{-st}\, \mathrm{d}t =\frac{r+s}{s}\ \tilde{G}_0(x, x_0,r+s)  \;
\end{equation}
where
\begin{equation}
   \tilde{G}_0(x,x_0,s) = \int_0^\infty  G_0(x,x_0,t) \, e^{-st}\, \mathrm{d}t = \dfrac{1}{\sqrt{4Ds}}\exp\left(-\sqrt{\dfrac{s}{D}}|x-x_0|\right) \;.
\end{equation}  
In arriving at the last equality, we used the explicit free propagator $G_0(x,x_0,t) = e^{-(x-x_0)^2/(4 D t)}/\sqrt{4 \pi D t}$. Substituting this result in Eq. (\ref{eqn:last_renewal_Laplace}), we obtain $\tilde{G}_r(x,x_0,s)$. What we need is actually $U_r(z,t) = \int_0^\infty G_r(x,-z,t)\,dx$. Taking the Laplace transform of this relation with respect to $t$ and using the explicit expression of $\tilde{G}_r(x,-z,s)$ we get
\begin{equation}
    \label{eqn:U_Brownian_Laplace}
    \tilde{U}_r(z,s) = \int_0^\infty  U_r(z,t) \, e^{-st}\, \mathrm{d}t =\frac{1}{2s}\exp\left[-\sqrt{\frac{r+s}{D}}z\right] \;.
\end{equation}
Inverting this Laplace transform (\ref{eqn:U_Brownian_Laplace}) with respect to $s$ using the convolution theorem, we obtain
\bea \label{Ur_t_diff}
U_r(z,t) = \int_{0}^{t} \frac{z}{4\sqrt{\pi D\tau^3}}\exp\left(-r\tau- \frac{z^2}{4D\tau} \right) \mathrm{d}\tau \;.
\eea
Note that while for finite $t$ it is hard to evaluate this integral explicitly, it simplifies in the large $t$ limit where it approaches a stationary form
\bea \label{Ur_stat}
U_r(z) = U_r(z,t \to \infty) = \frac{1}{2} e^{-\sqrt{\frac{r}{D}}\,z} \;.
\eea

Having obtained the central quantity $U_r(z,t)$ in Eq. (\ref{Ur_t_diff}) or equivalently its Laplace transform in 
Eq. (\ref{eqn:U_Brownian_Laplace}), we now discuss the annealed and the quenched cases separately.

\subsection{Annealed case}

As discussed before for general non-interacting particles, the current distribution $P_{\rm an}(Q,t)$ in the annealed case is Poissonian. This Poisson distribution is  fully characterized just by its mean $\mu_r(t) = \rho \int_0^{\infty} dz ~ U_r(z,t)$, where $U_r(z,t)$ 
is given in Eq. (\ref{Ur_t_diff}). To compute this integral, it is actually convenient to first consider its Laplace transform 
\begin{equation}
    \tilde{\mu}_r(s) = \int_0^\infty  \mu_r(t) \, e^{-st}\, \mathrm{d}t = \int_0^\infty \tilde U_r(z,s)\mathrm{d}z = \dfrac{\rho}{2s} \sqrt{\dfrac{D}{r+s}} \;,
\end{equation}
where in the last equality we used Eq. (\ref{eqn:U_Brownian_Laplace}). One can now invert this Laplace transform easily to obtain the exact mean current 
\begin{equation}
    \mu_r(t)= \frac{\rho}{2}\sqrt{\frac{D}{r}}\erf\left(\sqrt{rt}\right) \;, \label{mu_rBM}
\end{equation}
valid for all $t$. Thus in this case, the current distribution is exact at all time $t$. As discussed in Section \ref{sec:main}, the distribution becomes stationary at long time with mean $\mu_r(t \to \infty) = (\rho/2) \, \sqrt{D/r}$.

\subsection{Quenched case}

From Eq. (\ref{def_Ip}), we see that the basic ingredient needed to compute $P_{\rm qu}(Q,t)$ is also the function $U_r(z,t)$ already computed in Eqs. (\ref{eqn:U_Brownian_Laplace}) and (\ref{Ur_t_diff}). We start by computing the mean and the variance of $P_{\rm qu}(Q,t)$ given in Eqs. (\ref{mean_qu}) and (\ref{var_qu}). For the mean we obtain
\bea
    \label{average_flux_quenched}
    \langle Q\rangle_\mathrm{qu} &=\rho \int_{0}^{\infty} U_r(z, t)dz= \mu_r(t) = \frac{\rho}{2}\sqrt{\frac{D}{r}}\erf\left(\sqrt{rt}\right)
\eea    
where we used the expression for $\mu_r(t)$ from Eq. (\ref{mu_rBM}). Similarly, for the variance, we get     
\bea    
    \label{variance_flux_quenched}
    \sigma_{\mathrm{qu}}^{2}&=\left\langle Q^{2}\right\rangle_{\mathrm{qu}}-\langle Q\rangle_{\mathrm{qu}}^{2}=\rho \int_{0}^{\infty} U_r(z, t)(1-U_r(z, t)) d z \;,
\eea    
where $U_r(z,t)$ is given in Eq. (\ref{Ur_t_diff}). While it is difficult to perform the integral for $U_r(z,t)$ in Eq. (\ref{Ur_t_diff}), it turns out that the 
expression for the variance in Eq. (\ref{variance_flux_quenched}) simplifies when one performs 
the integral over $z$ first and then over $\tau$. This allows us to obtain an explicit formula for the variance
\bea \label{sigma_qu_diff}
\sigma_{\rm qu}^2 = \frac{\rho}{2} \, \sqrt{\frac{D}{r}}\, V(r\,t) \;,
\eea    
where the scaling function $V(z)$ is given by
\bea \label{Vofz}
V(z) = {\rm erf}(\sqrt{z}) - \frac{1}{4} \left[ \left( 1 - 2(2z+1)\,{\rm erfc}(\sqrt{z}) + (4z+1)\, {\rm erfc}(\sqrt{2z})  \right) + \frac{4}{\sqrt{\pi}} \sqrt{z}\,e^{-z} - \frac{2}{\sqrt{\pi}} \sqrt{2z} \,e^{-2z} \right] \;.
\eea    
This function $V(z)$ has the asymptotic behaviors
\bea \label{asympt_vz}
V(z) \approx
\begin{cases}
&\sqrt{\frac{2}{\pi}} \, \sqrt{z} + O(z^{3/2}) \quad, \quad z \to 0\\
& \\
& \frac{3}{4} + O(e^{-z}/\sqrt{z})  \quad, \quad \quad z \to \infty \;.
\end{cases}
\eea   
Thus $V(z)$ saturates to a constant $3/4$ as $z \to \infty$, indicating from Eq. (\ref{sigma_qu_diff}) that the variance approaches a constant as $t \to \infty$
\bea
\sigma_{\rm qu}^2 \to \frac{3\,\rho}{8} \, \sqrt{\frac{D}{r}}\;.
\eea
This is indeed the variance of the current in the stationary state. Indeed, in the stationary state, one can calculate all the cumulants of $P_{\rm qu}(Q,t \to \infty)$. To see this, we take the $t \to \infty$ limit of Eq. (\ref{def_Ip}). Substituting the stationary form $U_r(z, t \to \infty)$ in Eq. (\ref{Ur_stat}) and performing the integral in Eq. (\ref{def_Ip}), we get
\bea \label{Ip_infty}
I(p, t \to \infty) = - \rho \sqrt{\frac{D}{r}}\, {\rm Li}_2\left( \frac{1}{2}(1-e^{-p})\right) \;,
\eea
where ${\rm Li}_2(z) = \sum_{n\geq 1}z^n/n^2$ is the di-logarithm function, which is convergent for $|1-e^{-p}|<2$. For $p$ outside this range, one needs to analytically continue this formula (\ref{Ip_infty}) to obtain $I(p, t \to \infty)$. For small $p$, one can use this expression for $I(p, t \to \infty)$ in (\ref{Ip_infty}) 
and expand in powers of $p$ to compute all the cumulants of $P_{\rm qu}(Q, t \to \infty)$, i.e.,
\bea \label{cumul}
I(p, t \to \infty) = \sum_{n=1}^\infty \frac{(-p)^n\,\kappa_n}{n!} \;,
\eea
where $\kappa_n$ is the $n$-th cumulant of $P_{\rm qu}(Q, t \to \infty)$. Expanding Eq. (\ref{Ip_infty}) in powers of $p$ one can obtain all the cumulants in the stationary state as $\kappa_n = \rho \sqrt{r/D}\, a_n$, where the first $a_n$'s are given by
\bea \label{first_cumul}
a_1 = \frac{1}{2} \quad, \quad a_2 = \frac{3}{8} \quad, \quad a_3 = \frac{5}{24} \quad, \quad a_4 = \frac{1}{32} \quad,  \quad a_5 = - \frac{19}{240} \quad, \quad {\rm etc}. 
\eea
Note that the first two cumulants agree with the large $t$ limit of our exact formulae for the mean and the variance valid for all $t$, respectively in Eq. (\ref{average_flux_quenched}) and (\ref{sigma_qu_diff}). Interestingly, the fifth-th cumulant is negative. While we can compute the cumulants exactly in the stationary state, extracting $P_{\rm qu}(Q,t \to \infty)$ explicitly requires the knowledge of $I(p, t \to \infty)$ in the whole complex $p$-plane, which is more complicated and we do not pursue it further here.

So far, we have computed exactly the full time-dependent mean and the variance of the quenched current distribution and higher cumulants only in the stationary state. It turns out that one can also extract the behavior of the full distribution $P_{\rm qu}(Q,t)$ in the special scaling limit  
$Q\to+\infty$, $t \to +\infty$ but keeping the ratio $Q/t$ fixed. In this limit, we now show that it satisfies a large deviation behavior as announced in Eqs. (\ref{Pqu_diff}) and (\ref{large_deviation_function_psi_diffusion}).

This large $Q$ behavior can be extracted by analysing the Laplace transform in Eq. (\ref{Pqu2}). 
It turns out that, for this, we need to extract the behavior of $I(p,t)$ for large {\it negative} $p$. Analytically continuing the expression for $I(p,t)$ in Eq. (\ref{def_Ip}) to negative $p$, and using $v=-p$, we get 
\begin{equation} \label{def_Itilde}
    I(-v,t)\coloneqq \tilde{I}(v,t)\approx \rho \int_{0}^{+\infty} \mathrm{d}z \ln\left[1+e^v U_r(z,t) \right] \;,
\end{equation} 
 where we recall that the Laplace transform of $U_r(z,t)$ is given explicitly in Eq. (\ref{eqn:U_Brownian_Laplace}). Even though this Laplace transform can be inverted formally to obtain (\ref{Ur_t_diff}), it is useful to extract the asymptotic behaviors of $U_r(z,t)$ (for large $z$ and large $t$), directly from the Laplace transform. The reason for this is as follows. We will see later that for the RTP case, the 
  real space representation of $U_r(z,t)$ is much more complicated, while it is much easier in the Laplace space. Hence, working in the Laplace space is more convenient in both cases. The inverse Laplace transform $U_r(z,t)$ in Eq. (\ref{eqn:U_Brownian_Laplace}) can be expressed as a Bromwich integral in the complex $s$-plane
\bea \label{invert_U_diff}
U_r(z,t) = \int_{\Gamma} \frac{ds}{2 \pi i} \frac{e^{s\,t}}{2s}\, e^{-\sqrt{\frac{r+s}{D}}z} \;,
\eea
where $\Gamma$ runs along the vertical axis in the complex $s$-plane such that its real part is to the right of all the singularities of the integrand. We note that this integrand clearly has a simple pole at $s=0$. It also has a saddle point for large $t$ at $s=s^*$, where $s^*$ is obtained by minizing the argument of the exponential and is given by 
\bea \label{sp_diff}
s^* = - r + \frac{z^2}{4 D\, t^2} \;.
\eea 
Now two situations can occur: 
\begin{itemize}
\item[(i)] {$s^*<0$, i.e., $z < \sqrt{4 D \,r}\, t$: in this case, the dominant contribution to the integral comes from the pole at $s=0$ and one simply gets
\bea \label{pole_diff}
U_r(z,t) \approx \frac{1}{2}\, e^{-\sqrt{\frac{r}{D}}\,z} \;.
\eea
\item[(ii)] {$s^*>0$, i.e., $z > \sqrt{4 D \,r}\, t$: in this case the saddle occurs to the right of the pole on the real axis. Hence one can deform the Bromwich contour to pass through the saddle to pick up the leading contribution. Evaluating this saddle point and ignoring pre-exponential factors, we get
\bea \label{saddle_diff}
U_r(z,t) \approx e^{-t \left( r + \frac{z^2}{4D\,t^2} \right)} \;.
\eea

}

}
\end{itemize} 
\begin{figure}[t]
\includegraphics[width=\linewidth]{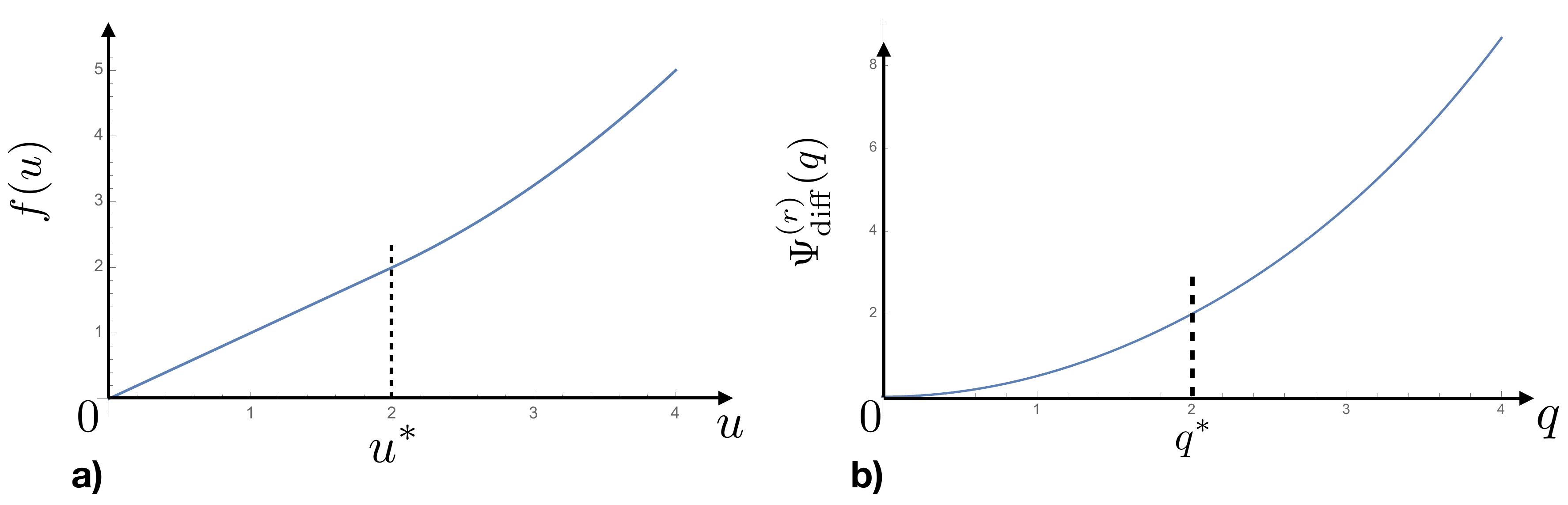}
\caption{{\bf a)}: Plot of the function $f(u)$ vs $u$ in Eq. (\ref{large_deviation_diffusion_U}). With the parameter choices $D=1, r=1$, we have $u^* = \sqrt{4D\,r} = 2$ shown by the vertical dashed line where the function $f(u)$ and its first derivative are continuous, but the second derivative is discontinuous. {\bf b)} Plot of the rate function $\Psi_{\rm diff}^{(r)}(q)$ vs $q$ as given in Eq. (\ref{large_deviation_function_psi_diffusion.2}). The critical value $q^* = 2 \sqrt{D/r}\,\rho = 2$, for the choice $D=1, r=1$ and $\rho = 1$, is marked by the vertical dashed line.} \label{Fig_func_ann}
\end{figure}

Combining these two behaviors, (\ref{pole_diff}) and (\ref{saddle_diff}) one can express $U_r(z,t)$ in a large deviation form
 \begin{equation}
\label{large_deviation_diffusion_U}
    U_r(z,t)\thicksim e^{-t\,f\left(\frac{z}{t}\right)},\quad \text{where}\quad f(u)=
\begin{dcases}
    \sqrt{\dfrac{r}{D}}u &\text{if } u<u^*=\sqrt{4Dr} \;,\\
    r+\dfrac{u^2}{4D} \quad &\text{if } u>u^*=\sqrt{4Dr} \;.
\end{dcases}
\end{equation}
The rate function $f(u)$ and its first derivative is continuous at $u = u^* = \sqrt{4D\,r}$ (see Fig. \ref{Fig_func_ann} a)). However, the second derivative is discontinuous. This is similar to the second order transition found in the position distribution of a resetting Brownian motion in Ref. \cite{MSS15a}. Substituting this large deviation form (\ref{large_deviation_diffusion_U}) in Eq. (\ref{def_Itilde}) we get 
\begin{equation} \label{Itilde_largedev}
    I(-v,t)\coloneqq \tilde{I}(v,t)  \approx \rho \int_{0}^{+\infty} \mathrm{d}z \ln\left[1+(e^v-1)\,e^{-t\,f\left(\frac{z}{t}\right)} \right]\approx \rho \int_{0}^{+\infty} \mathrm{d}z \ln\left[1+e^ve^{-t\,f\left(\frac{z}{t}\right)} \right] \;,
\end{equation}
where, in the last approximation, we used $v \gg 1$. To evaluate this integral, we use the following trick: we first take the partial derivative of $\tilde I(v,t)$ with respect to $v$, then we compute the integral and finally integrate back with respect to $v$. The partial derivative gives
\begin{align}  \label{Fermi}
    \frac{\partial \tilde{I}(v,t)}{\partial v} \approx \rho\int_{0}^{+\infty}\mathrm{d}z \frac{1}{1+e^{-[v-tf(z/t)]}} = \rho \, t\int_{0}^{+\infty}\mathrm{d}x \frac{1}{1+e^{-t[v/t-f(x)]}}.
\end{align}
Let us define $x^*$ such that 
\be \label{xstar}
v=t\,f(x^*) \;,
\ee
where $f(x)$ is given in Eq. (\ref{large_deviation_diffusion_U}). Hence Eq. (\ref{Fermi}) reads 
 \begin{align}  \label{Fermi2}
    \frac{\partial \tilde{I}(v,t)}{\partial v} =  \rho \, t\int_{0}^{+\infty}\mathrm{d}x \frac{1}{1+e^{-t[f(x^*)-f(x)]}} \;.
\end{align}
In the limit of large $t$, we now recognise that the integrand in Eq. (\ref{Fermi2}) is like a Fermi function (at ``an inverse temperature'' $t$). Hence for large $t$ (zero-temperature limit), one can replace the Fermi-function by a step function, which takes value $1$ for $x<x^*$ and $0$ for $x>x^*$. Hence the integral over $x$ in (\ref{Fermi2}) gets cut-off at $x^*$ and we get to leading order for large $t$
\begin{equation}
\frac{\partial \tilde{I}(v,t)}{\partial v} \approx \rho t\,x^* = \rho \,t\,f^{-1}\left(\frac{v}{t}\right)
\end{equation}
where $f^{-1}$ is  the inverse function of $f$. It can be easily read off by inverting Eq. (\ref{large_deviation_diffusion_U}) explicitly  
\begin{equation} \label{finv}
f^{-1}(y) =\begin{cases}
\sqrt{\dfrac{D}{r}}y & \text{ for } \quad y<2r \;,\\
\sqrt{4D(y-r)} & \text{ for } \quad y>2r \;.
\end{cases}
\end{equation}
Finally integrating back in $v$, with $v/t$ fixed, we get to leading order for large $t$
\begin{equation}
\tilde{I}(v,t)=\rho t \int_0^{v}f^{-1}\left(\frac{y}{t}\right)\mathrm{d}y=\rho t^2 \int_0^{\frac{v}{t}}f^{-1}(y)\mathrm{d}y \;.
\end{equation}
Using the explicit form of $f^{-1}(y)$ from Eq. (\ref{finv}), we arrive at
\bea \label{scalingI}
\tilde{I}(v,t) \approx -\rho t^2 \, \tilde{\phi}\left(\frac{v}{t}\right) 
\eea
with 
\begin{equation}
\label{tilde_phi_diffusion}
\tilde{\phi}(y)=\begin{cases}
-\sqrt{\dfrac{D}{r}}\dfrac{y^2}{2} & \text{ for }y<2r\\ \\
-\dfrac{2}{3}\sqrt{Dr^3}-\dfrac{4}{3}\sqrt{Dr^3}\left(\dfrac{y}{r}-1\right)^{3/2} & \text{ for }y>2r \;.
\end{cases}
\end{equation}
Substituting this scaling form (\ref{scalingI}) of $\tilde I(v,t)$ on the right hand side of Eq. (\ref{Pqu2}), 
we then get
\bea \label{exppQ}
\langle e^{-p Q}\rangle \approx \int_{0}^\infty e^{-p\,Q} P_{\rm qu}(Q,t) dQ \approx \exp{\left( -\rho t^2 \, \tilde{\phi}\left(\frac{v}{t}\right) \right)} \;.
\eea
We then see that, for consistency, one needs the following large deviation form for $P_\mathrm{qu}(Q,t)$ 
\begin{equation} \label{large_dev_ansatz}
P_\mathrm{qu}(Q,t)\thicksim \exp\left[-\rho t^2\psi\left(\frac{Q}{\rho t}\right) \right] \;,
\end{equation}
where $\psi(q)$ is yet to be determined. Indeed, substituting this form on the left hand side of \eqref{Pqu2} one gets
\begin{align}
\langle e^{-pQ}\rangle &= \int_{0}^{+\infty}e^{-pQ}P_\mathrm{qu}(Q,t) \mathrm{d}Q \notag \sim \int_{0}^{+\infty}\exp\left\lbrace -\rho t^2\left[ \dfrac{Q}{\rho t^2}p+\psi\left(\dfrac{Q}{\rho t}\right)\right] \right\rbrace \mathrm{d}Q
\end{align}
setting $w=\dfrac{Q}{\rho t}$ and $p=-v$:
\begin{equation} \label{saddle_point}
\langle e^{-pQ}\rangle \thicksim \int_{0}^{+\infty} \exp\left\lbrace -\rho t^2\left[\psi(w)-\dfrac{v}{t}w \right] \right\rbrace \mathrm{d}w \thicksim \exp\left\{-\rho t^2 \left(\min_{w>0} [\psi(w) - \frac{v}{t} w] \right)\right\} \;,
\end{equation}
where we used a saddle-point approximation, valid for large $t$. Comparing Eq. (\ref{saddle_point}) and (\ref{exppQ}), we arrive at the Legendre transform
\begin{equation} \label{legendre}
\min_{w>0}\left[\psi(w)-y\,w\right]=\tilde{\phi}(y)\quad\Rightarrow\quad \psi(w)=\max_{y}\;[\tilde{\phi}(y)+w\,y]\;.
\end{equation}
Substituting the form of $\tilde{\phi}(y)$ given in Eq. (\ref{tilde_phi_diffusion}) and maximising we get
\bea \label{psi_w}
\psi(w) =
\begin{cases}
&\frac{1}{2} \sqrt{\frac{r}{D}}\, w^2 \quad, \quad \quad \quad \quad \quad \quad \; w < \sqrt{4\,r \,D}\;, \\
& \\
& - \frac{2}{3} \sqrt{D\,r^3} + r\,w + \frac{w^3}{12D} \;, \; \quad w > \sqrt{4\,r \,D} \;.
\end{cases}
\eea
By inserting this form on the right hand side of Eq. (\ref{large_dev_ansatz}) gives us the required large deviation form of $P_{\rm qu}(Q,t)$. It has a slightly nicer form in terms of the dimensionless variable $q = Q/(r\,t)$. In terms of $q$, Eq. (\ref{large_dev_ansatz}) then reads
\bea
\label{large_deviation_form_P_qu}
 P_\mathrm{qu}(Q,t) \thicksim \exp\left\lbrace-r^2t^2\Psi_\mathrm{diff}^{(r)}\left(q=\frac{Q}{r\,t}\right)\right\rbrace \;,
\eea 
where 
\begin{equation}
\label{large_deviation_function_psi_diffusion.2}
\Psi_\mathrm{diff}^{(r)}(q)=\begin{cases}
\dfrac{q^2}{q^*}\quad & \text{ for } q<q^*=2\sqrt{\dfrac{D}{r}}\rho\\
-\dfrac{q^*}{3}+q+\dfrac{q^3}{3{q^*}^2} & \text{ for } q>q^*=2\sqrt{\dfrac{D}{r}}\rho \;.
\end{cases}
\end{equation}
This rate function is plotted in Fig. \ref{Fig_func_ann} b). The non-analytic behavior at $q=q^*$ where the third derivative of the rate function $\Psi^{(r)}_\mathrm{diff}(q)$ is discontinuous represents a third order transition as mentioned before. In Section \ref{sec:num}, we verify this analytical prediction in numerical simulations using importance sampling.Interestingly, the extremely fast decay of the quenched probability in Eq. (\ref{large_deviation_form_P_qu}), namely $P_\mathrm{qu}(Q,t)  \sim e^{-t^2}$, can be understood from the following simple physical argument. We recall that $Q$ denotes the number of particles to the right of the origin at time $t$. Hence a configuration where $Q \sim t$ originates typically from trajectories where an initial $\rho \, t$ number of particles to the left of the origin does not undergo any reset up to time $t$ (otherwise, they will not be able to cross the origin easily). The probability that a single particle does not undergo any reset up to time $t$ is simply $\sim e^{-r\,t}$. Since the particles are independent, the probability for such a configuration then is of order $[e^{-r\,t}]^{\rho\,t} \sim e^{- t^2}$, up to some exponent which is indeed the large deviation function $\Psi_{\rm diff}^{(r)}(q)$.

Finally, we note that in the very large time limit when $r\,t \gg 1$, the scaled variable $q \to 0$ and using the first line of Eq. (\ref{large_deviation_function_psi_diffusion.2}) in (\ref{large_deviation_form_P_qu}), we get a Gaussian tail of the steady state $P_{\rm qu}(Q,t \to \infty)$ 
\bea \label{Gauss_tail}
P_{\rm qu}(Q \gg 1,t \to \infty) \sim e^{- \frac{1}{2\rho}\sqrt{\frac{r}{D}}Q^2}  \;.
\eea
Note that this Gaussian tail is not so easy to derive directly from the exact cumulant generating function of  $P_{\rm qu}(Q,t \to \infty)$ in Eq. (\ref{Ip_infty}).

\section{Run and tumble particles with stochastic resetting}\label{sec:RTP}

In the case of RTP's, the trajectory of a single particle is specified by both the position $x_i(t)$ and the velocity $v_i(t) = \pm v_0$ at
time $t$. We assume that the initial positions of the non-interacting RTPs are chosen independently from a uniform step initial condition
(as in the diffusive case above). Furthermore, we assume that the initial velocities are chosen independently as $\pm v_0$ with equal probability. After each resetting event, the position $x_i(t) \to x_i$ and the velocity $v_i(t) \to \pm v_0$, i.e., the velocity gets re-randomized after each
resetting. Let us first denote by $G_r(x,x_0,t)$ the propagator of this RTP process with resetting where both the initial and the final velocities are summed over, i.e., it represents the ``marginalised'' position propagator. One can then write an exact renewal equation, similar to Eq. (\ref{eqn:last_renewal_eq}) for the diffusive case~\cite{EM18,EMS20}
\begin{equation}
    \label{eqn:last_renewal_eqRTP}
    G_r(x,x_0,t) = e^{-rt}G_0(x,x_0,t) +r\int_{0}^{t} e^{-r\tau} G_0(x,x_0, \tau) \mathrm{d}\tau \;,
\end{equation}
where $G_0(x,x_0,\tau)$ now represents the position propagator of an RTP, starting from random initial velocities $\pm v_0$ with equal probability, whose expression is known \cite{Weiss}
\begin{equation}\label{ac-fullpd3}
G_0(x,x_0,t) = \frac{e^{-\gamma t}}{2}\left\{ \delta(x-x_0-v_0t) + \delta(x-x_0+v_0t) +\frac{\gamma}{v_0}\left[I_0(\omega) + \frac{\gamma t I_1(\omega)}{\rho} \right] \theta(v_0t-|x-x_0|) \right\},
\end{equation}
where $\omega$ is given by
\begin{eqnarray}\label{def_omega}
\omega = \frac{\gamma}{v_0}\sqrt{v_0^2t^2 -(x-x_0)^2} \;.
\end{eqnarray} 
Once again, the central quantity is the function 
\bea \label{U_RTP}
U_r(z,t) = \int_0^\infty G_r(x,-z,t) \, dx \;,
\eea
where $G_r(x,-z,t)$ satisfies Eq. (\ref{eqn:last_renewal_eqRTP}). Integrating the renewal equation (\ref{eqn:last_renewal_eqRTP}) over $x$ and using the definition of $U_r(z,t)$ above, we get a renewal equation for $U_r(z,t)$
 \begin{equation} \label{Ur_RTP}
    U_r(z,t) = e^{-rt} U_0(z,t) +r\int_{0}^{t} e^{-r\tau} U_0(z,\tau) \mathrm{d}\tau \;,
\end{equation}
where $U_0(z,\tau) = \int_0^\infty G_0(x,-z,\tau)\,dx$ with $G_0$ given in Eq. (\ref{ac-fullpd3}). For later purposes, it is useful to note that the Laplace transform of $U_r(z,t)$ with respect to $t$ in Eq. (\ref{Ur_RTP}) has a compact form, 
\bea \label{Ur_Laplace}
\tilde U_r(z,s) = \int_0^\infty e^{-s\,t}\, U_r(z,t) = \frac{r+s}{s} \, \tilde U_0(z,r+s) \;.
\eea
The Laplace transform of $U_0(z,t)$ can be obtained exactly from Eq. (\ref{ac-fullpd3}) -- see \cite{Banerjee_2020} -- and has a compact expression
\bea \label{tildeU_zs}
\tilde U_0(z,s) = \frac{1}{2s} \, \exp{\left(-\frac{\sqrt{s(s+2\gamma)}}{v_0}\,z \right)} \;.
\eea
Using this result in Eq. (\ref{Ur_Laplace}) we get
\bea \label{Ur_Laplace.2}
\tilde U_r(z,s) = \frac{1}{2\,s} \exp{\left( -\frac{\sqrt{(r+s)(r+s+2\gamma)}}{v_0}\,z\right)} \;.
\eea
This is the central result we need to discuss the annealed and the quenched cases below.

\subsection{Annealed case}

As for an non-interacting particle systems, the current distribution in the annealed case is Poissonian, and is fully characterized
by its mean 
\bea  \label{mur_rtp}
\mu_r(t) = \rho \int_0^\infty U_r(z,t)\, dz  \;.
\eea
Thus, we just need to compute this mean $\mu_r(t)$ to characterise the full distribution $P_{\rm an}(Q,t)$. As in the diffusive case, it turns out to be convenient to consider first its Laplace transform with respect to $t$
\bea\label{mUr_Laplace}
\tilde \mu_r(s) = \int_0^\infty \mu_r(t) e^{-s t} \, dt = \rho \int_0^\infty \tilde U_r(z,s)\, dz = \frac{\rho\,v_0}{2s\sqrt{(r+s)(r+s+2\gamma)}} \;,
\eea
where in the last equality we used (\ref{Ur_Laplace.2}). In the small $s$ limit the right hand side of (\ref{mUr_Laplace}) behaves as $\sim 1/s$, indicating that $\mu_r(t)$ approaches a constant in the long time limit and is given by
\bea  \label{mur_inf}
\mu_r(t \to \infty) =   \frac{\rho \, v_0}{2\sqrt{r(r+2\gamma)}} \;.
\eea 
In fact, $\mu_r(t)$ for finite $t$ can be obtained by inverting formally the Laplace transform in (\ref{mUr_Laplace}) using the convolution theorem, leading to 
\begin{equation} \label{mur_RTP2}
    \mu_r(t) = e^{-rt}\mu_0(t) +r\int_0^t e^{-r\tau}\mu_0(\tau) \mathrm{d}\tau \;,
\end{equation}
where 
\begin{equation} \label{mu0_RTP}
    \mu_0(t) = \dfrac{1}{2} \rho v_0\, t\, e^{-\gamma t} [I_0(\gamma t) +I_1(\gamma t)] \;,
\end{equation}
was already computed in Ref.~\cite{Banerjee_2020}.

\subsection{The quenched case}\label{subsection_diffusive_dynamics_quenched}

\begin{figure}[t]
\includegraphics[width = \linewidth]{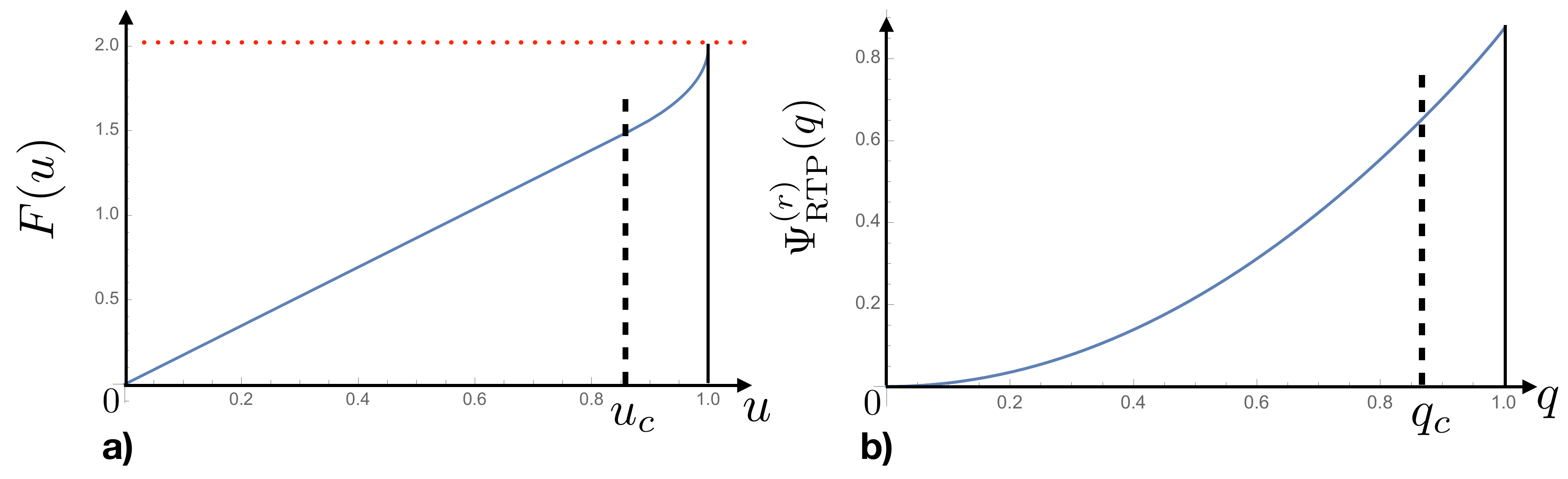}
\caption{{\bf a)}: Plot of the function $F(u)$ vs $u$ in Eq. (\ref{large_deviation_RTP_U}) for $0 \leq u \leq 1$. With the parameter choices $\gamma=1, r=1$, we have $u_c = \sqrt{r(r+2\gamma)}/(r+\gamma) = \sqrt{3}/2=0.866025\ldots$ shown by the vertical dashed line where the function $F(u)$ and its first derivative are continuous, but the second derivative is discontinuous. The maximal value of the function $F(u)$ is attained at $u=1$ where $F(1) = \gamma + r = 2$ as shown by the horizontal red dotted line. {\bf b)}: Plot of the rate function $\Psi_{\rm RTP}^{(r)}(q)$ vs $q$ with $0\leq q \leq 1$, as given in Eq. (\ref{large_dev_RTP4}). The critical value $q_c =  \sqrt{r(r+2\gamma)}/(r+\gamma) = \sqrt{3}/2=0.866025\ldots$ is marked by the vertical dashed line.}\label{Fig_func_qu}
\end{figure}

As in the annealed case, the central information for the quenched case is also encoded in the function $U_r(z,t)$ whose Laplace transform is given in Eq. (\ref{Ur_Laplace.2}). The mean and the variance are again given by Eqs. (\ref{mean_qu}) and (\ref{var_qu}). For the mean we get the same answer as in the annealed case 
\begin{align} \label{av_qu_RTP}
\langle Q\rangle_\mathrm{qu} &= \mu_r(t) 
\end{align}
where the Laplace transform of $\mu_r(t)$ is given in Eqs. (\ref{mur_RTP2}) and (\ref{mu0_RTP}). Since we do not have any explicit expression for $U_r(z,t)$, the expression for the variance is a bit cumbersome to obtain explicitly. Rather, we focus on the large deviation behavior of $P_{\rm qu}(Q,t)$ as in the resetting Brownian motion. In order to extract this behavior from Eq. (\ref{Pqu2}) and (\ref{def_Ip}), we need the asymptotic behavior of $U_r(z,t)$ when $z$ and $t$ are both large. To find this behavior, we follow the same analysis as in the diffusive case, namely we 
formally invert the exact Laplace transform in Eq. (\ref{Ur_Laplace.2}) as
\bea \label{inv_Laplace}
U_r(z,t) = \int_{\Gamma} \frac{ds}{2\pi i} \frac{1}{2\,s} \exp{\left[t \left(s -{\sqrt{(r+s)(r+s+2\gamma)}}\,\frac{z}{v_0\,t}\right)\right]} \;,
\eea
where $\Gamma$ represents the vertical Bromwich contour in the complex $s$-plane, passing to the right of all the singularities. Once again, this integrand has a pole at $s=0$ and a saddle at $s=s^*$ (obtained by minimising the argument of the exponential). After straightforward algebra, we find that there are actually two possible values of $s^*$, as solutions of a quadratic equation. One of these roots is always negative and do not contribute to the large time analysis of the integral. The explicit expression of the largest one is given by
\bea \label{s_star_RTP}
s^* = - r  + \gamma \left( \frac{1}{\sqrt{1 - \left( \frac{z}{v_0t}\right)^2}} - 1 \right)  \;.
\eea 
As in the diffusive case, there are two possibilities
\begin{itemize}
\item[(i)] {$s^*<0$, i.e., $z < \frac{\sqrt{r(r+2\gamma)}}{r + \gamma}\,v_0\, t$, where : in this case, the dominant contribution to the integral comes from the pole at $s=0$ and one simply gets
\bea \label{pole_RTP}
U_r(z,t) \approx \frac{1}{2}\, e^{-\sqrt{r(r+2 \gamma)}\,\frac{z}{v_0}} \;.
\eea
\item[(ii)] {$s^*>0$, i.e., $z > \frac{\sqrt{r(r+2\gamma)}}{r + \gamma}\,v_0\, t$: in this case the saddle occurs to the right of the pole on the real axis. Hence one can deform the Bromwich contour to pass through the saddle to pick up the leading contribution. Evaluating this saddle point and ignoring pre-exponential factors, we get after a few steps of algebra
\bea \label{saddle_RTP}
U_r(z,t) \approx \exp{\left[-t \left( r + \gamma  - \gamma \,\sqrt{1 - \left(\frac{z}{v_0\,t}\right)^2} \right)\right]}\; \theta(v_0\,t - z) \;.
\eea
Physically, it is clear that $z \leq v_0\,t$ since, if the absolute value of the initial position of a particle is $z  > v_0\,t$ it can not contribute to the 
current $Q$ up to time $t$, which is simply the number of particles to the right of the origin at time $t$. 
}
}
\end{itemize}

Combining these two results (\ref{pole_RTP}) and (\ref{saddle_RTP}), the function $U_r(z,t)$ for large $z$ and large $t$ but with $z/(v_0t)$ fixed can be expressed in a large deviation form
\begin{equation}
\label{large_deviation_RTP_U}
    U_r(z,t)\thicksim e^{-t\,F\left(\frac{z}{v_0\,t}\right)},\quad \text{where}\quad F(u)=
\begin{dcases}
  \sqrt{r(r+2 \gamma)}\,u &\text{if } u<u_c = \frac{\sqrt{r(r+2\gamma)}}{(r+\gamma)} \;,\\
   r+\gamma - \gamma \,\sqrt{1 - u^2}  \quad &\text{if } u_c<u<1 \;.
\end{dcases}
\end{equation}
A plot of this function $F(u)$ is given in Fig. \ref{Fig_func_qu} a). We note that the function $F(u)$ is supported over $0\leq u \leq 1$. As $u \to 1$, $F(u)$ approaches its maximal value $F(1) = r+\gamma$. Also at $u=u_c$, the function $F(u)$ and its first derivative $F'(u)$ are continuous, while the second derivative is discontinuous (as in the diffusive case in Eq. (\ref{large_deviation_diffusion_U}) for $f(u)$).

We now substitute this expression of $U_r(z,t)$ from Eq. (\ref{large_deviation_RTP_U}) into Eq. (\ref{def_Ip}) and follow a similar analysis as was done for the resetting Brownian motion. As in that case, to analyse the large $Q$ behavior of $P_{\rm qu}(Q,t)$ in Eq. (\ref{Pqu2}), we need to analyse the behavior of $I(p,t)$ in Eq. (\ref{def_Ip}) in the limit $p \to - \infty$. Setting $v = -p$ with $v>0$, we write
\begin{equation} \label{I_p_RTP}
I(v,t)\coloneqq \tilde{I}(v,t) =\rho \int_{0}^{v_0\,t} \mathrm{d}z \ln\left[1+(e^v-1)U_r(z,t) \right] \approx \rho \int_{0}^{v_0\,t} \mathrm{d}z \ln\left[1+e^vU_r(z,t) \right].
\end{equation}

As before, the trick is to take the partial derivative with respect to $v$, to compute the integral and finally to integrate back in $v$. Substituting the form of $U_r(z,t)$ given in (\ref{large_deviation_RTP_U}) in (\ref{I_p_RTP}) and taking the 
derivative with respect to $v$, we get
\begin{align}  \label{Fermi_RTP}
    \frac{\partial \tilde{I}(v,t)}{\partial v} \approx \rho\int_{0}^{v_0\,t}\mathrm{d}z \frac{1}{1+e^{-[v-t\,F(z/(v_0\,t))]}} = \rho \, v_0\,t\int_{0}^{1}\mathrm{d}x \frac{1}{1+e^{-t[v/t-F(x)]}}.
\end{align}
Let us define $x^*$ such that 
\be \label{xstar_RTP}
v=t\,F(x^*) \;,
\ee
where $F(x)$ is given in Eq. (\ref{large_deviation_RTP_U}). Hence Eq. (\ref{Fermi}) reads 
 \begin{align}  \label{Fermi2_RTP}
    \frac{\partial \tilde{I}(v,t)}{\partial v} =  \rho \,v_0 \,  t\int_{0}^{1}\mathrm{d}x \frac{1}{1+e^{-t[F(x^*)-F(x)]}} \;.
\end{align}
Since $F(x)$ is a monotonically increasing function, for $x>x^*$, the integrand essentially vanishes in the large $t$ limit. In contrast, for $x<x^*$, the integrand approaches $1$ as $t \to \infty$. Thus, once again, the integrand behaves as a Fermi function at ``an inverse temperature'' $t$. Hence, cutting the integral at $x=x^*$, we get for large $t$
\begin{equation} \label{deriv}
\frac{\partial \tilde{I}(v,t)}{\partial v} \approx \rho \, v_0 \, t\,x^* = \rho \,v_0\,t\,F^{-1}\left(\frac{v}{t}\right) \;,
\end{equation}
where $F^{-1}$ is  the inverse function of $F$. This inverse function can be easily extracted from the explicit form of $F(u)$ in Eq. (\ref{large_deviation_RTP_U}). One gets
\bea \label{Finv}
F^{-1}(y) = 
\begin{cases}
&\frac{y}{\sqrt{r(r+2 \gamma)}} \quad, \quad \quad \quad \;\, {\rm for} \quad y < y_c = \frac{r(r+2\gamma)}{r+\gamma} \\
& \\
& \sqrt{1 - \frac{(\gamma+r-y)^2}{\gamma^2}}  \quad, \quad {\rm for} \quad y_c < y < \gamma + r \;.
\end{cases}
\eea
Note that the upper limit $y = \gamma + r$  comes from the fact that the maximal value of $F(u)$ is $\gamma + r$ [see Fig. \ref{Fig_func_qu} a)]. If $v/t > \gamma+r$, then the integrand is always $1$ as $t \to \infty$, since $v/t > F(x)$ for all $x \in [0,1]$.

Integrating this relation (\ref{deriv}), and considering the two regimes $v/t < \gamma + r$ and $v/t > \gamma + r$, one gets to leading order for large $t$
\bea \label{integr1}
\tilde I(v,t) \approx - \rho\,v_0\, t^2 \, \tilde \Phi \left( \frac{v}{t}\right) \;,
\eea
where 
\bea \label{phi_RTP}
\tilde \Phi(y) = 
\begin{cases}
&- \int_0^{y} F^{-1}(y') \, dy'  \quad, \quad \hspace*{2.8cm} y < \gamma + r \;,\\
& \\
& - \int_0^{\gamma + r} F^{-1}(y') \, dy' - (y - (\gamma + r)) \quad, \quad y > \gamma + r \;.
\end{cases}
\eea

Let us now relate the large deviation behavior of $P_{\rm qu}(Q,t)$ to this function $\tilde \Phi(y)$. Substituting this form (\ref{integr1}) in Eq. (\ref{Pqu2}) with $p=-v$, one gets
\bea \label{large_dev_RTP}
\int_0^\infty  P_{\rm qu}(Q,t)\, e^{v\,Q} \, dQ \sim e^{- \rho\,v_0\, t^2 \, \tilde{\Phi}\left( \frac{v}{t}\right)} \;;.
\eea 
This form suggests the following large deviation form for $P_{\rm qu}(Q,t)$
\bea \label{large_dev_RTP2}
P_{\rm qu}(Q,t) \sim e^{- \rho\, v_0 \, t^2 \, \Psi \left( \frac{Q}{\rho\, v_0\,t}\right)} \;.
\eea
Substituting this form on the left hand side of Eq. (\ref{large_dev_RTP}), performing a saddle point analysis for large $t$, we get (as in the diffusive case)
\bea \label{legendre_RTP}
\min_{0\leq q\leq 1}\left[\Psi(q)-y\,q\right]=\tilde{\Phi}(y)\quad\Rightarrow\quad \Psi(q)=\max_{y}\; [\tilde{\Phi}(y)+q\,y]\ \;.
\eea
Our next goal is to compute $\tilde \Phi(y)$ and then use Eq. (\ref{legendre_RTP}) to compute $\Psi(q)$. It turns out that the regime $y > \gamma + r$ in the second line of Eq. (\ref{phi_RTP}) does not contribute to the computation of $\Psi(q)$ in Eq. (\ref{legendre_RTP}), as can be verified a posteriori. We therefore restrict to the region $y < \gamma + r$, as given in the first line of Eq. (\ref{phi_RTP}). To compute $\tilde \Phi(y)$, we first insert 
$F^{-1}(y')$ from (\ref{Finv}) in (\ref{phi_RTP}) and integrate. This gives
\bea \label{expl_tildephi}
\tilde \Phi(y) = 
\begin{cases}
& - \dfrac{y^2}{2 \sqrt{r(r+2 \gamma)}} \quad, \quad \hspace*{4.6cm} y < y_c = \dfrac{r(r+2 \gamma)}{r+\gamma} \;, \\
& \\
& - \dfrac{y_c^2}{2 \sqrt{r(r+2 \gamma)}} - {\mathlarger{\int}}_{\hspace*{-0.15cm}y_c}^y \sqrt{1 - \dfrac{(\gamma+r-y')^2}{\gamma^2}}\, dy'  \quad, \quad y_c< y < \gamma + r \;.
\end{cases}
\eea
Substituting this in Eq. (\ref{legendre_RTP}) and maximising with respect to $y$, one gets an explicit expression for $\Psi(q)$. We omit the details here, since they are quite straightforward but involve lengthy algebra. We just remark that when one maximises with respect to $y$ in Eq. (\ref{legendre_RTP}) for a fixed $0\leq q \leq 1$, it turns out that the value $y^*$ that maximises $\tilde \Phi(y) + q\,y$ is always less than $\gamma + r$. In fact, as $q$ approaches its maximal value $1$, the value of $y^*$ approaches $\gamma + r$. This justifies, a posteriori, the restriction to the range $y< \gamma +r$ in Eq. (\ref{expl_tildephi}).

Let us then just quote the final result here. In fact, it is more natural to express the large deviation form in terms of a dimensionless rate function. Hence we re-write Eq. (\ref{large_dev_RTP2}) as
\bea \label{large_dev_RTP3}
P_{\rm qu}(Q,t) \sim e^{- \rho\, v_0\, \gamma \, t^2 \, \Psi_{\rm RTP}^{(r)} \left( \frac{Q}{\rho\, v_0\,t}\right)} \;,
\eea
with the dimensionless rate function $\Psi_{\rm RTP}^{(r)}(q) = \Psi(q)/\gamma$. In terms of the dimensionless ratio $\alpha = r/\gamma$, 
the rate function $\Psi_{\rm RTP}^{(r)}(q)$ can then be expressed as
\bea \label{large_dev_RTP4}
\Psi_{\rm RTP}^{(r)}(q) = 
\begin{cases}
&\dfrac{q_c}{2\sqrt{1 - q_c^2}}\, q^2 \quad, \quad \hspace*{7.8cm} 0< q < q_c = \dfrac{\sqrt{\alpha(\alpha+2)}}{\alpha+1}\;. \\
& \\
& -\dfrac{q_c}{2 \sqrt{1- q_c^2}} + \dfrac{q}{\sqrt{1-q_c^2}} - \dfrac{1}{2} q\,\sqrt{1 - q^2} + \dfrac{1}{2} {\rm sin}^{-1}(q_c) - \dfrac{1}{2} {\rm sin}^{-1}(q) \quad, \quad q_c < q \leq 1 \;.
\end{cases}
\eea
A plot of this function for $q \in [0,1]$ is given in Fig. \ref{Fig_func_qu} b). The function $\Psi_{\rm RTP}^{(r)}(q)$ and its two first derivatives are continuous at $q=q_c$. However the third derivative is discontinuous at $q=q_c$. Hence, this rate function for the RTP with resetting also exhibits a third order phase transition, as in the case of resetting Brownian motion. 

Let us comment on few other properties of this rate function $\Psi_{\rm RTP}^{(r)}(q)$. 
\begin{itemize}
\item{} From the large deviation form (\ref{large_dev_RTP3}), we see that in the large time limit, with $Q$ fixed, the current distribution $P_{\rm qu}(Q,t)$ approaches a time-independent stationary form 
\bea \label{stationary_qu_RTP}
P_{\rm qu}(Q,t)  \underset{t \to \infty}{\longrightarrow}  \exp{\left(- \frac{1}{2 \rho} \sqrt{\frac{2\gamma\,r}{v_0^2}}\,Q^2\right)} \;.
\eea
In the diffusive limit, when $\gamma \to \infty$, $v_0 \to \infty$, with $v_0^2/(2 \gamma) = D$ fixed, this result coincides exactly with Eq. (\ref{P_qu_diff_stat}) for the resetting Brownian motion. 

\item{} When the resetting rate $r \to 0$, the ratio $\alpha = r/\gamma \to 0$ and hence $q_c \to 0$. In this case, the rate function $\Psi_{\rm RTP}^{(r=0)}$ is entirely given by the second line of Eq. (\ref{large_dev_RTP4}) for the full range $0 \leq q \leq 1$ and it reads  
\bea \label{large_dev_RTP5}
\Psi_{\rm RTP}^{(r=0)}(q)  = q - \frac{q}{2}\sqrt{1-q^2} - \frac{1}{2}\, {\rm sin}^{-1}(q) \;.
\eea
This indeed coincides with the result of Ref. \cite{Banerjee_2020}, which we recalled in the introduction in Eq. (\ref{rtp-ldf-model}). Note that, using the identity ${\rm sin}^{-1}(q) = {\rm sin}^{-1}\left[ \sqrt{\frac{1-\sqrt{1-q^2}}{2}} \right]$ (for $0\leq q \leq 1$), one sees that (\ref{rtp-ldf-model}) and (\ref{large_dev_RTP5}) are identical.

\item{} Finally, in the limit $q \to 1$, it follows from Eq. (\ref{large_dev_RTP4}) that 
\bea \label{psi_q1}
\Psi_{\rm RTP}^{(r)}(q=1) = - \frac{q_c}{2\sqrt{1-q_c^2}} + \frac{1}{\sqrt{1-q_c^2}} + \frac{1}{2} {\rm sin}^{-1}(q_c) - \frac{\pi}{4} \;,
\eea
where we recall that $q_c = \sqrt{\alpha(\alpha+2)}/(\alpha + 1)$ with $\alpha = r/\gamma$. Using this result in Eq. (\ref{large_dev_RTP3}), one predicts that the probability of the rare event that the current $Q$ takes its 
maximally allowed value, $Q = \rho\, v_0\, t$, decays extremely rapidly with time as 
\bea
P_{\rm qu}(Q = \rho \,v_0 \, t,t ) \sim \exp{\left(- \theta \, t^2\right)} \quad, \quad \theta = \rho\, v_0 \, \gamma\, \left[ - \frac{q_c}{2\sqrt{1-q_c^2}} + \frac{1}{\sqrt{1-q_c^2}} + \frac{1}{2} {\rm sin}^{-1}(q_c) - \frac{\pi}{4}\right] \;.
\eea
In the absence of resetting, i.e., when $r \to 0$, we get $\alpha \to 0$ and hence $q_c \to 0$. Consequently, $\theta \to  \rho\, v_0 \, \gamma (1 - \pi/4)$, reproducing the result of Ref. \cite{Banerjee_2020}. As the resetting rate $r$ increases, the probability that the current $Q$ achieves its maximal value becomes more unlikely and hence one would expect the exponent $\theta$ to be a monotonically increasing function of $r$, which is indeed the case.

\end{itemize}

\section{Numerical simulations}\label{sec:num}

In the previous sections, we have computed analytically the current distribution $P(Q,t)$, both the annealed and the quenched versions,
for two different models, but both with stochastic resetting. In the first model, the underlying dynamics of each particle is an independent Brownian motion with stochastic resetting, while in the second model, it is the run-and-tumble dynamics with positon resetting and simultaneous velocity randomisation. In the annealed case the current distribution is Poissonian in both models, only the mean differs in the two models. In contrast, in the quenched case, the distributions are more complex and exhibit a nontrivial large deviation behavior summarized in Eqs. (\ref{Pqu_diff})-(\ref{large_deviation_function_psi_diffusion}) for the diffusive case and in Eqs. (\ref{large_dev_qu_intro})-(\ref{Psi_r_intro}) for the RTP. Note that in both cases, these probabilities are extremely tiny as they decay as $\sim e^{-t^2}$ for large $t$. Therefore computing these rate functions numerically is a formidable challenge and cannot be achieved by standard Monte-Carlo sampling which will always miss such extremely rare events. In this section, we use a specialised importance sampling method, designed precisely to capture such tiny probabilities. We present here the results for the resetting Brownian motion. The same method can be extended to the case of resetting RTP also, but we will not repeat this simulation here and restrict ourselves only to the diffusive case below.

In order to obtain the distributions over a large range of the support, down to extremely small
probabilities, we use a numerical large-deviation algorithm~\cite{DZ10}.
In general, for complex interacting systems, one needs a sophisticated Markov-chain Monte Carlo
simulations to sample distributions in several regions of their support, via biased sampling of random numbers~\cite{Hart02,Hart14,Hart15,Hart18}.
However, here, where the particles do not interact with each other, a much simpler approach can be used.

\subsection{Algorithm}

\begin{figure}[t]
\begin{center}
    \includegraphics[width=0.6\textwidth]{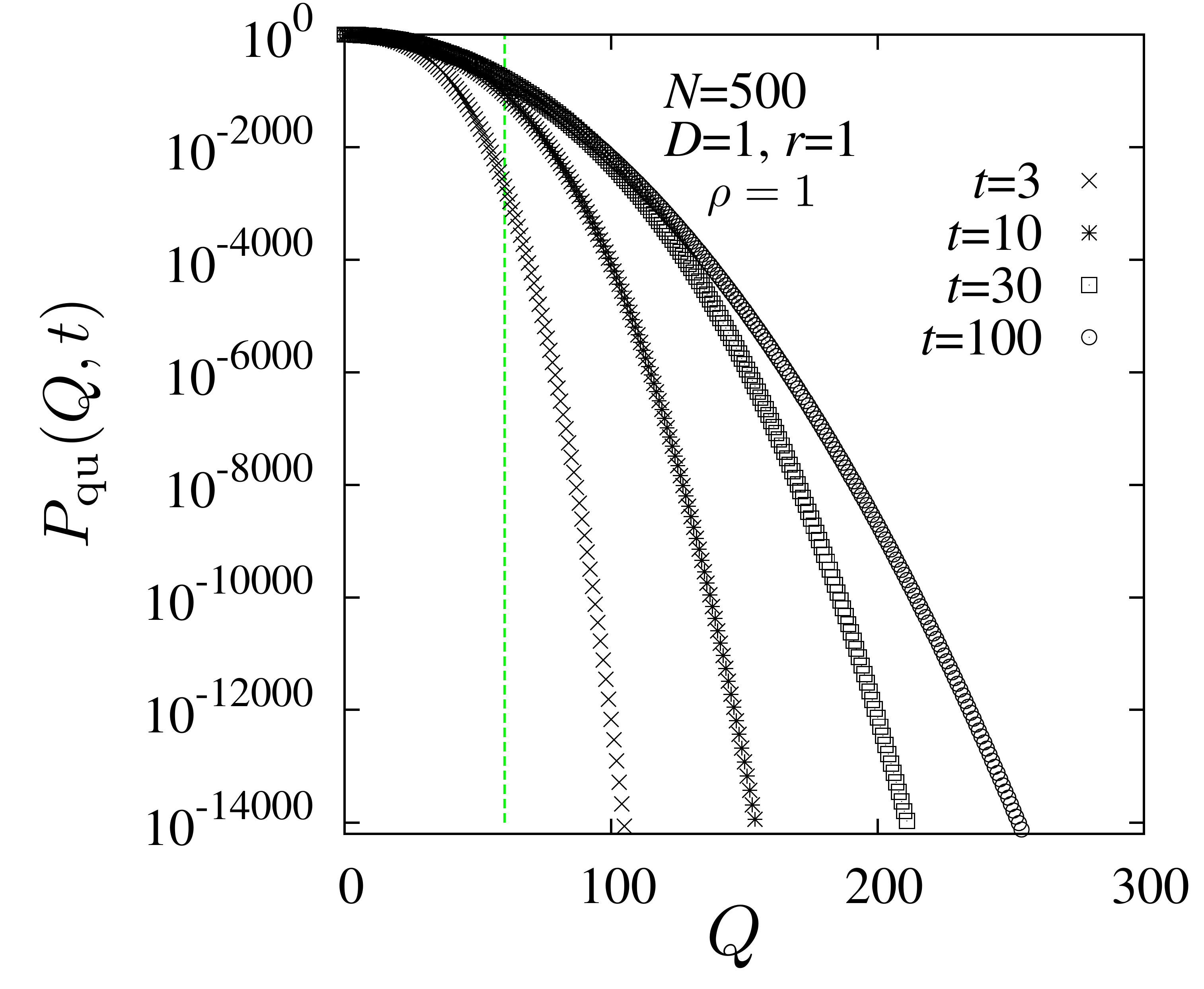}
\end{center}
\caption{\label{fig:PQ_qu} Distribution $P_{\rm qu}(Q,t)$ of the current for the quenched case for $N=500$ particles, parameters $D=1$, $r=1$, $\rho=1$ and four different times $t=3,10,30,$ and $100$. The vertical green dashed line corresponds to the critical value $Q^*(t)=rt\,q^*$ for $t=30$ with $q^* = 2 \sqrt{D/r}\,\rho$, see Eq.~(\ref{large_deviation_function_psi_diffusion}). For $Q \ll Q^*(t)$, one expects the distribution to be stationary, i.e., independent of $t$, and its logarithm with a quadratic form, as in Eq. (\ref{P_qu_diff_stat}).}
\end{figure}

We consider $N$ particles labelled by $i=1,2, \cdots, N$.Their initial locations are marked as $x_i \leq 0$. Each particle $i$ diffuses and resets to its initial position $x_i$ with a rate $r$. The probability $p^+_i$ that particle $i$ contributes to the flux $Q$ at time $t$ is the probability that its position at time $t$ is in the positive half-space, i.e.,
\begin{equation}
p^+_i\equiv U_r(-x_i,t) =\int_0^{+\infty} dx\, G_r(x,x_i,t) \,,
\end{equation}
 where the propagator $G_r(x,x_i,t)$ is given in Eq.~(\ref{eqn:last_renewal_eq}) with
\begin{equation} \label{G0_num}
G_0(x,x_i,\tau)=\frac{1}{\sqrt{4\pi D \tau}}
\exp\left( -\frac{(x-x_i)^2}{4D\tau}\right) \,.
\end{equation}
Since we consider starting positions in the negative range, the probabilities are often extremely small. For a simpler numerical treatment, we shift the argument of the Gaussian by a suitable particle-dependent value $+f_i$ and compensate this by a factor $e^{-f_i}$ in front of the integral. We have chosen  $f_i=x_i^2/(4Dt)$ which was convenient. Using Eq. (\ref{G0_num}) in (\ref{eqn:last_renewal_eq}), with this additional bias, 
we get
\begin{eqnarray}
\label{eq:p_plus}
    p^+_i & \equiv p^+(x_i) =  e^{-f_i} & \int_0^{+\infty} dx  \left[ 
    \frac{e^{-rt}}{\sqrt{4\pi D t}}
\exp\left( -\frac{(x-x_i)^2}{4Dt}+f_i\right)  \right. \\
& & \quad \quad \quad \quad +
r \int_0^t d\tau\, \left.
\frac{e^{-r\tau}}{\sqrt{4\pi D \tau}}
\exp\left( -\frac{(x-x_i)^2}{4D\tau}+f_i\right) \right] \;. \nonumber
\end{eqnarray}

For the quenched case, the dominant contribution to $P_{\rm qu}(Q,t)$ comes from a typical configuration where the initial positions are equispaces with separation $1/\rho$, i.e., $x_1=0$, $x_2=-1/\rho$, $\ldots, x_N=-(N-1)/\rho$ \cite{Banerjee_2020}. We have numerically determined the values $p^+_i$ $(i=1,\ldots,N)$ by calculating the integrals $\int_0^{+\infty} dx$ and $\int_0^t d\tau$ in Eq. (\ref{eq:p_plus}) by using the adaptive stepsize \verb!gsl_integration_qag()! integration function of the GNU scientific library \cite{gsl2006}. The resulting probabilities, which include the $e^{-f_i}$ factors, can be very small, thus we have used for the further processing a custom-made datatype for very small or large numbers. {We have used in our C implementation two {\tt double} numbers to represent large numbers, one for the mantissa and one for the exponent. %One can also use special libraries like the GNU \emph{MP Bignum} \cite{mp_bignum} library.}.

For the annealed case, the probabilities $p^+_i$ do not depend on a single starting position any more, but are just averages of Eq. (\ref{eq:p_plus}) over the initial positions, which are uniformly distributed in $[-L,0]$, where $L=N/\rho$. This
yields
\begin{equation}
    p^+_i = \frac 1 L \int_{-L}^0 dz\, p^+(z)\,
\end{equation}
for all particles $i$. We have performed this integral by using the same GSL function. Note that since the integral is dominated by starting positions
near $z=0$, no shift of the Gaussian is necessary, i.e., $f=0$. Thus, the annealed probability is much larger than the quenched-case $p^+(x_i)$ for most starting positions $x_i$.

\begin{figure}[t]
\begin{center}
    \includegraphics[width=0.6\textwidth]{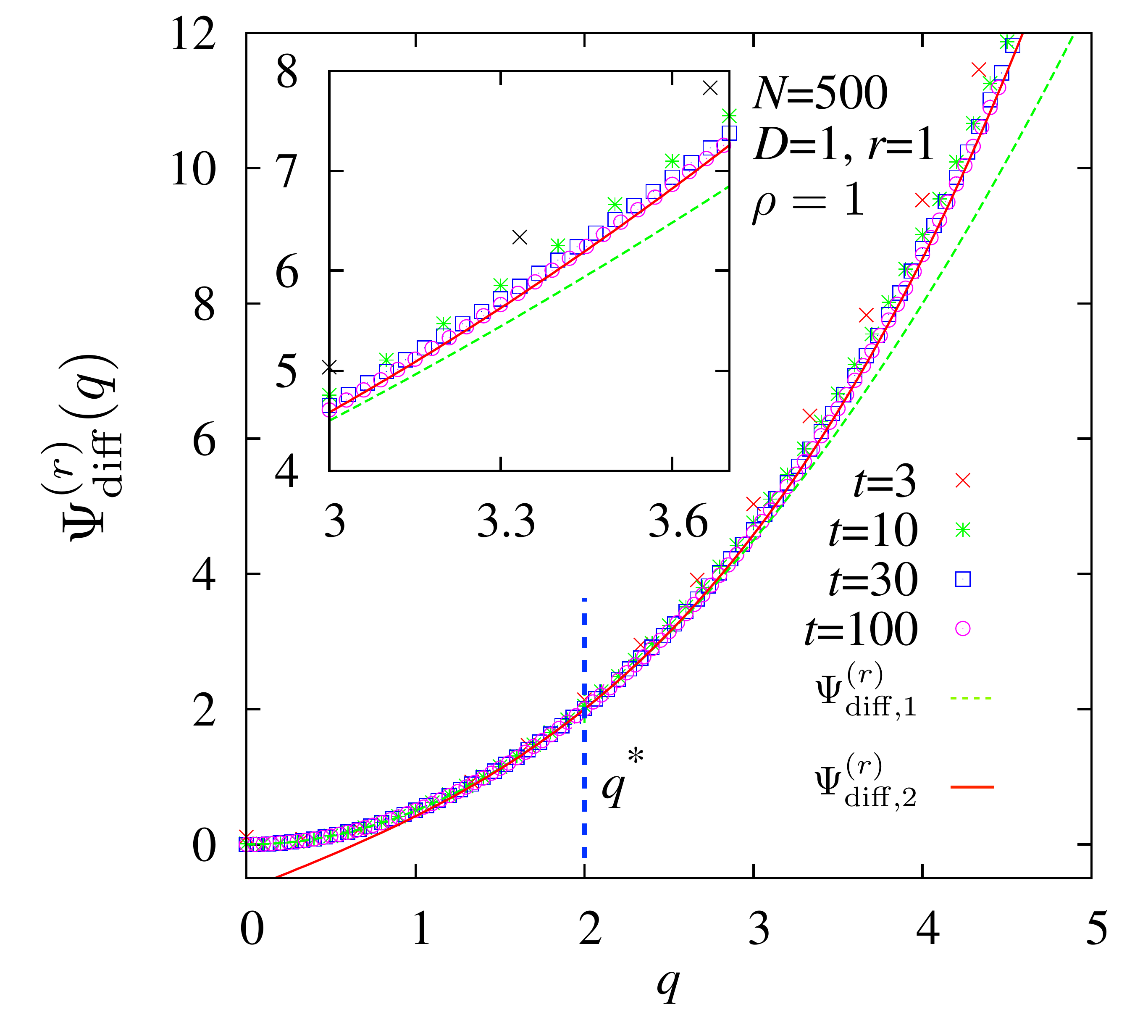}
\end{center}
\caption{\label{fig:psi_qu} The symbols show the numerical 
rate function $\Psi^{(r)}_{\rm diff}(q)$ associated with the distribution of the 
scaled current $q=Q/(rt)$ for the quenched case for $N=500$ particles, with parameters $D=1$, $r=1$, $\rho =1$ and for four different times $t=3,10,30,$ and $100$. The dashed green line represents the branch $\Psi^{(r)}_{{\rm diff},1}(q)$, given in the first line of Eq. (\ref{large_deviation_function_psi_diffusion}). The solid red line represents the branch $\Psi^{(r)}_{{\rm diff},2}(q)$, given in the second line of Eq. (\ref{large_deviation_function_psi_diffusion}). They intersect at $q=q^* = 2\sqrt{D/r} \rho = 2$ shown by the vertical dashed blue line. The inset shows a blow-up of the region $q=[3.0,3.7]$, where we see the convergence to the branch $\Psi^{(r)}_{{\rm diff},2}(q)$ (the solid red line) better.}
\end{figure}

Now, the probability $P(Q,t)$ for a specific value of the flux $Q$, is the sum over all combinations of particle positions at time $t$, where exactly $Q$ particles have a positive position, which appears with probability $p^+_i$, and $N-Q$ particles have a negative position, which appears with probability $1-p^+_i$, respectively. For the annealed case, this is simply the binomial distribution, which could be in principle be directly evaluated, but we use, for simplicity, the same approach as for the quenched case. For the quenched case, one would have to enumerate all exponentially many possible combinations of a number $Q$ of particles having a positive coordinate, and the other ones a negative coordinate. Since this is not feasible, we use sampling.

For simple sampling, to generate one sample value of $Q$, one would iterate over all particles, generating a configuration $I=(I_1,\ldots,I_N)$. For each particle with probability $p_i^+$ the particles exhibits a positive coordinate, denoted by $I_i=1$, which generates a contribution of $1$ to $Q=Q(I)=\sum_i I_i$. With probability $1-p_i^+$ the particles contributes $I_i=0$. By repeating this assignment to $I$ many times, one will sample many values of $Q$, which results in a histogram which can be normalized. But this allows only to estimate the high probability part of $P(Q)$.

For this reason, we used a biased sampling approach. We intended to sample configurations $I$ exhibiting a flux $Q(I)$, where an addition bias $e^{\beta Q(I)}$ is applied, with a variable parameter $\beta$, which acts as a negative inverse temperature. For large values of $\beta$, the resulting underlying distribution $P_{\beta}(Q)\sim P(Q)e^{\beta Q} $ will be shifted to  higher values of $Q$. The basic idea is to obtain the distribution for several parameter values of $\beta$ and obtain the true distribution $P(Q)$ by combining the measured distributions $P_{\beta}(Q)$, see below. 

The mentioned bias means for a single particle that it obtains a bias $e^{\beta}$ for contributing to the flux, i.e., having a positive position, while there is no bias for a negative position. Therefore, we sample the state $I_i$ of particle $i$ with the distribution
\begin{eqnarray}
    \tilde P(I_i)=\delta(I_i-1) \frac{p_i^+ e^{\beta}}{p_i^+e^{\beta}+(1-p_i^+)}
    +\delta(I_i) \frac{1-p_i^+}{p_i^+e^{\beta}+(1-p_i^+)}\,.
\end{eqnarray}
Trivially, the state of each particle can be sampled directly from the given two probabilities. Thus, for each iteration one independent sample configuration $I$ with value $Q(I)$ for the biased ensemble will be obtained. By taking a decent number of samples, one can easily get high precision histograms estimating the typical, i.e., high probability part of $P_\beta(Q)$, respectively. We have
sampled $10^6$ independent configurations for each value of $\beta$.

\begin{figure}[t]
\begin{center}
    \includegraphics[width=0.6\textwidth]{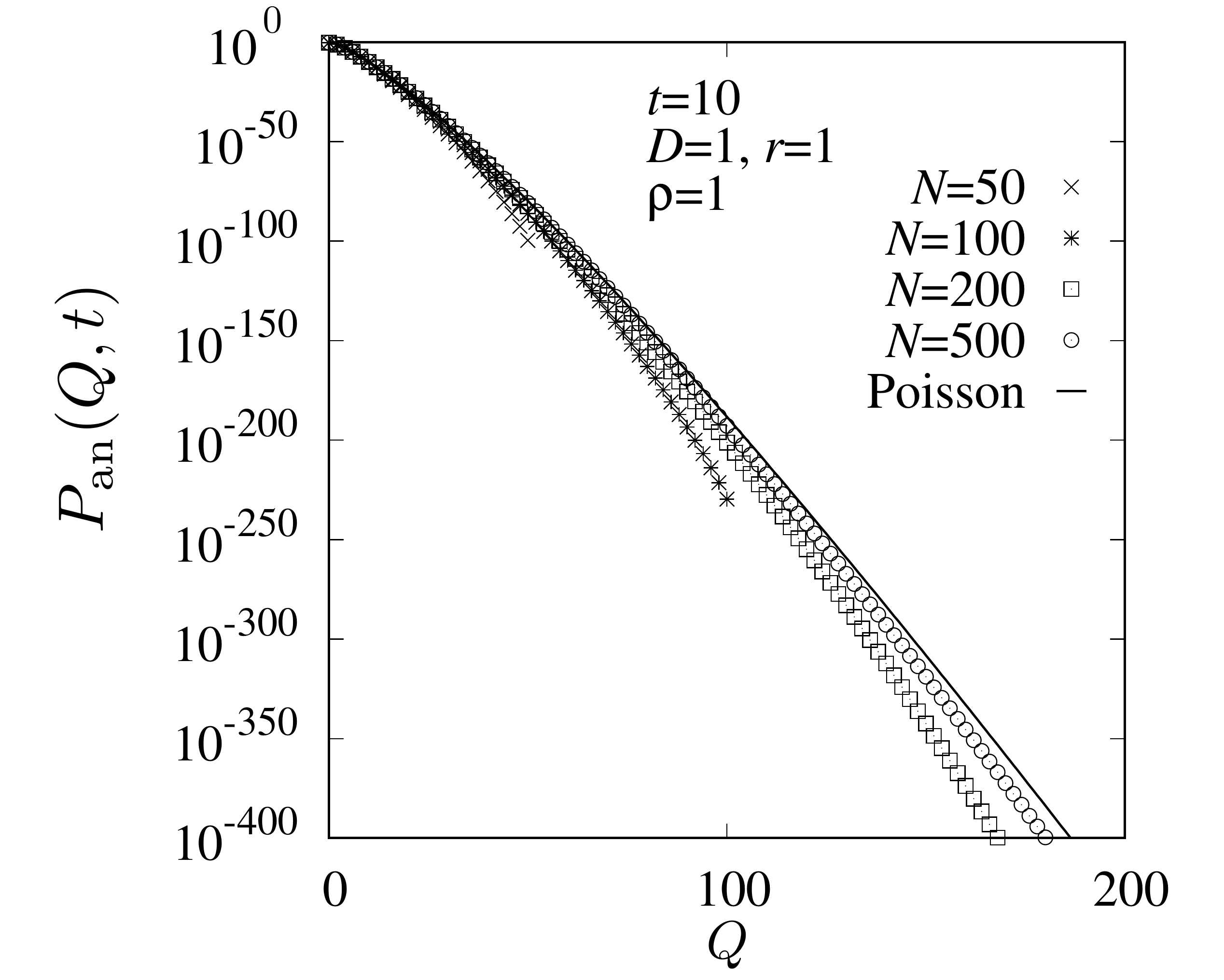}
\end{center}
\caption{\label{fig:PQ_an} Current distribution $P_{\rm an}(Q,t)$ for the annealed case at time $t=10$, with parameters $D=1$, $r=1$, $\rho = 1$ and for four different number
of particles $N = 50,100,200,$ and $500$. As $N$ increases, the distribution approaches an $N$-independent limiting Poissonian form with mean $\mu_r(t) \approx 0.5$.}
\end{figure}

Within this biased sampling, each configuration $I$ will appear with the following probability:
\begin{eqnarray}
    \tilde P_\beta(I) & = & \prod_{i|I_i=1} 
    \frac{p_i^+ e^{\beta}}{p_i^+e^{\beta}+(1-p_i^+)}
    \prod_{i|I_i=0} \frac{1-p_i^+}{p_i^+e^{\beta}+(1-p_i^+)} \nonumber \\
    & = & \frac {e^{\beta Q(I)}}{Z} \prod_{i|I_i=1} p_i^+ \prod_{i|I_i=1} (1-p_i^+)\\
    & = & \frac {e^{\beta Q(I)}}{Z}  P_0(I)
\end{eqnarray}
where the normalization is given by $Z=\prod_i (p_i^+e^{\beta}+(1-p_i^+))$,
and the sampling for $\beta=0$ is equal to the unbiased original ensemble. Thus,
for the distribution of $Q$, where one sums over all configurations $I$
which exhibit the flux $Q$, one obtains, for all values of $\beta$:
\begin{equation}
    P(Q) \equiv P_0(Q)= Z e^{-\beta Q}P_\beta(Q)\,.
\end{equation}
Hence, to obtain the estimate for the distribution $P(Q)$, for each value of $Q$, we took, for simplicity,
the normalized histogram $h_{\beta^*}(Q)$ at that value $\beta^*$ which exhibits the highest histogram value, i.e., the best statistics, and simply calculated   $Z e^{-\beta^* Q}h_{\beta^*}(Q)$ as estimate for $P(Q)$. Typically, we needed a number $K$ of different values of $\beta$ which was in between 100 und 300 to sufficiently cover a large range of the support of $P(Q)$. Here, we used uniformly distributed values $\beta_k= k \Delta \beta$ ($k=0,\ldots,K-1$).
Depending on the case we considered, we used either 
$\Delta\beta=3$ or $\Delta\beta = 5$.

\subsection{Results}

For simulations, we chose the parameters $D=1$, $r=1$ and used $N=500$ particles with $\rho=1$. For the quenched case we have obtained $P_{\rm qu}(Q,t)$ for four different times $t=3$, $t=10$, $t=30$ and $t=100$. The resulting distributions are shown in
Fig.~\ref{fig:PQ_qu}. The distributions can be sampled down to very small probabilities like $10^{-14000}$ with high statistical accuracy. One observes
that with increasing times, larger values of $Q$ become more probable and 
the distributions approach more and more a common shape.

To allow for a comparison with the result given in
Eq.~(\ref{large_deviation_function_psi_diffusion}), we plot the rate
function $\Psi_{\rm diff}^{(r)}(q)=-\log(P_{\rm qu}(Q,t))/(r^2t^2)$ as a function of $q=Q/(rt)$ and compare
with the two branches $\Psi^{(r)}_{{\rm diff},1}(q)$ (for $q<q^*)$ and $\Psi^{(r)}_{{\rm diff},2}(q)$ (for $q>q^*$) given respectively in the first and the second line of Eq. (\ref{large_deviation_function_psi_diffusion}).
The result is shown in Fig.~\ref{fig:psi_qu}. For $q<q^*$, the result follows
for all times the analytical result $\Psi^{(r)}_{{\rm diff},1}(q)$ very well. For large values
$q>q^*$ a fast convergence towards the analytical result  $\Psi^{(r)}_{{\rm diff},1}(q)$ can be well observed with increasing time $t$.

%\begin{figure}
%\begin{center}
%    \includegraphics[width=0.9\textwidth]{Px_qu.pdf}
%\end{center}
%\caption{\label{fig:Px_qu} The symbols show the distribution of particle
%distances from the restting position for parameters $D=1$, $r=1$ and time $t=10$.
%The line shows the analytical result for the stationary case $t\to\infty$. The vertical lines
%indicate $\pm Q^*_{t=10}$.}
%\end{figure}

The result for the annealed current distribution $P_{\rm an}(Q,t)$ is shown in Fig.~\ref{fig:PQ_an} for $t=10$ and the parameter values $D=1$,  $r=1$ and $\rho = 1$. Since the annealed probability $p^+$ is much larger than most probabilities $p_i^+$, the
resulting probabilities of the current are also much larger. Still, the distribution stretches down to very small probabilities such as $10^{-400}$. With increasing number $N$ of particles a convergence to a limiting distribution, independent of $N$, is visible. This $N$-independent limiting distribution is Poissonian with mean $\mu_r(t) = (\rho/2) \sqrt{D/r}\,{\rm erf}(\sqrt{r\,t})$ as given in Eq. (\ref{eqn:mu_Brownian}). For $t=10$ and the given parameters above, $\mu_{r=1}(t=10) = (1/2) \,{\rm erf}(\sqrt{10})\approx 0.5$.

 \section{Conclusion}\label{sec:conclusion}

In this paper, our goal was to study to the effects of resetting on the distribution $P(Q,t)$ of the integrated particle current $Q$ up to time $t$ through the origin in a one-dimensional system. We studied the setting where the particles are non-interacting and are initially distributed on the left of the origin, with a uniform density $\rho$. We chose two different dynamics for the particles: (i) Brownian dynamics with stochastic resetting at rate $r$ and (ii) run and tumble dynamics with a persistence time $\gamma^{-1}$ and subjected to stochastic resetting with rate $r$. We studied both the annealed and the quenched current distributions. In both models, and for both annealed and quenched cases, the main effect of resetting is to induce a stationary limit to the current distribution at long times. 
It was previously known that the position distribution of a single particle under stochastic resetting becomes stationary at long times, but here we show a similar effect for the distribution of the current, which itself is a  dynamical observable.  

One of our main findings is that, in the presence of stochastic resetting, the approach to the stationary state of the current distribution in the annealed and the quenched cases are drastically different for both models. In the annealed case, the distribution $P_{\rm an}(Q,t)$ is Poissonian at all times and the whole distribution approaches its stationary limit at late times uniformly for all $Q$. In contrast, the quenched current distribution $P_{\rm qu}(Q,t)$ is highly non-Poissonian at all times $t$. Moreover, we showed that the approach to the stationary state in the quenched case is highly non-uniform in $Q$. More precisely,  
as time increases, we showed that there is a critical value $Q_{\rm crit}(t)$ that increases linearly with $t$ such that, for $Q < Q_{\rm crit}(t)$, the quenched distribution $P_{\rm qu}(Q,t)$ attains its stationary form, while it remains time-dependent for $Q > Q_{\rm crit}(t)$. 
This critical value $Q_{\rm crit}(t)$ thus separates the steady state from the transient regime. On this scale, when $Q \sim Q_{\rm crit}(t)$, we show that the quenched distribution admits an unusual large deviation form in both models. We have computed the corresponding rate functions analytically in both models. When $Q$ crosses the critical value $Q_{\rm crit}(t)$, we found that the associated rate functions undergo  a third order phase transition, such that the rate function and its first two derivatives are continuous, while the third derivative is discontinuous. Measuring numerically such a rate function presents a technical challenge. Here, using an importance sampling algorithm that is able to access probabilities as small as $10^{-14000}$, we were able to compute the rate function for the Brownian case with resetting. We found an excellent agreement with our analytical predictions.

Such third order phase transitions have been found previously in many different contexts, e.g., in the distribution of the largest eigenvalue of random matrices~\cite{MS14}, Yang-Mills gauge theory~\cite{GW80,Wadia80,DK93,FMS11}, Coulomb gases~\cite{CFLV17,Dhar17}
 and also in the height distribution of the $1+1$-dimensional Kardar-Parisi-Zhang fluctuating interface model at late times \cite{KPZ1,KPZ2,KD19}. In all these examples, the underlying system is strongly interacting. Hence, it is interesting that a similar third order transition occurs in such a seemingly simple noninteracting system.   
Let us also make an interesting technical remark here. In the quenched case, for both models, while the Legendre transform of the rate function has a second-order singularity at the critical point, the rate function itself has a third-order singularity. It would be interesting to investigate the generality of this mechanism.

There are many interesting directions in which the present work can be extended. For example, it would be interesting to see whether this third-order phase transition in the quenched case also occurs for other dynamics with resetting, beyond the Brownian and run and tumble particles studied here. Another interesting question is whether this transition persists in the presence of interactions between particles or for correlated initial conditions~\cite{BJC22}.

%\bibliography{Bibiography.bib}

\end{document}